%% file: main.tex
\documentclass[a4paper,11pt]{article}
\usepackage{aaskaiid}
\usepackage{orcidlink}
\usepackage{xcolor}
\usepackage{enumitem}
\usepackage{bm}
\usepackage{hyperref}
\usepackage{comment}
\usepackage{subcaption}

\def\Q{\mathcal{Q}}
\def\mpci{{\rm Mpc}^{-1}}
\def\CII{{\rm [CII]}--158}
\def\CI{{\rm [CI]}--371}

\def\NII{{\rm [NII]}--205}
\def\CO{{\rm CO}(1--0)}
\def\B19{\texttt{B19}}
\def\HI{HI}

\title{Cosmology with Multi-Wavelength Line Intensity Mapping Synergies in the SKAO Era}
\ShortTitle{Multi-Wavelength LIM Synergies with SKAO}

\author[1,2]{Debanjan Sarkar \orcidlink{0000-0001-5763-2541}}
\ShortName{Sarkar et al.} 
\author[3]{Suman Majumdar \orcidlink{0000-0001-5948-6920}}
\author[4]{Chandra Shekhar Murmu \orcidlink{0000-0002-1818-5440}}
\author[3]{Leon Noble \orcidlink{0009-0004-3138-1130}}
\author[3]{Manas Mohit Dosibhatla \orcidlink{0000-0003-4229-2972}}
\author[3]{Vishrut Pandya \orcidlink{0009-0009-6595-014X}}
\author[5]{José Luis Bernal \orcidlink{0000-0002-0961-4653}}
\author[3]{Yashrajsinh Mahida \orcidlink{0009-0000-1796-797X}}
\author[6]{Mohd Kamran \orcidlink{0000-0002-0107-9844}}
\author[7]{Caroline S. Heneka \orcidlink{0000-0001-8883-0583}}
\author[3]{Anshuman Tripathi \orcidlink{0000-0002-5091-9950}}
\author[3]{Samit Kumar Pal \orcidlink{0000-0002-2271-4165}}
\author[6]{Gabriella De Lucia \orcidlink{0000-0002-6220-9104}} 
\author[6]{Fabio Fontanot \orcidlink{0000-0003-4744-0188}}
\author[8]{Matteo Viel \orcidlink{0000-0002-2642-5707}}
\author[9,10]{Marta Spinelli \orcidlink{0000-0003-0148-3254}}
\author[3]{Parth Kothari \orcidlink{}}
\author[11]{Sarah Libanore \orcidlink{0000-0002-2284-9190}}
\author[12,13]{Abinash Kumar Shaw \orcidlink{0000-0002-6123-4383}}
\author[14]{Martin Sahlen \orcidlink{0000-0003-0973-4804}}
\author[15,16,10]{José Fonseca \orcidlink{0000-0003-0549-1614}}
\author[3]{Abhirup Datta \orcidlink{0000-0002-5333-1095}}
\author[17,18,19]{Ziad Sakr \orcidlink{0000-0002-4823-3757}}
\author[10]{Wenkai Hu \orcidlink{0000-0002-3108-5591}}

\affiliation[1]{Department of Physics and Trottier Space Institute, McGill University, QC H3A 2T8, Canada}
\affiliation[2]{Ciela---Montreal Institute for Astrophysical Data Analysis and Machine Learning, QC H2V0B3, Canada}
\affiliation[3]{Department of Astronomy, Astrophysics \& Space Engineering, Indian Institute of Technology Indore, Indore 453552, India}
\affiliation[4]{Astrophysics Research Centre of the Open University (ARCO) and Department of Natural Sciences, The Open University of Israel,
1 University Road, PO Box 808, Ra’anana 4353701, Israel}
\affiliation[5]{Instituto de Física de Cantabria (IFCA), CSIC-Univ. de Cantabria, Avda. de los Castros s/n, E-39005 Santander, Spain}
\affiliation[6]{INAF-Osservatorio Astronomico di Trieste, Via G. B. Tiepolo 11,
34143 Trieste, Italy}
\affiliation[7]{Institut f\"ur Theoretische Physik, Universit\"at Heidelberg, Philosophenweg 16, 69120 Heidelberg, Germany}
\affiliation[8]{SISSA, International School for Advanced Studies, Via Bonomea
265, 34136 Trieste TS, Italy}
\affiliation[9]{Observatoire de la Côte d’Azur, Laboratoire Lagrange, Bd de l’Observatoire, CS 34229, 06304 Nice cedex 4, France}
\affiliation[10]{Department of Physics and Astronomy, University of the Western Cape, Robert Sobukwe Road, Cape Town 7535, South Africa}
\affiliation[11]{Department of Physics, Ben-Gurion University of the Negev, Be’er Sheva 84105, Israel}
\affiliation[12]{Department of Computer Science, University of Nevada Las Vegas, 4505 S. Maryland Pkwy., Las Vegas, NV 89154, USA}
\affiliation[13]{Max-Planck-Institut für Astrophysik, Garching D-85748, Germany}
\affiliation[14]{Department of Physics and Astronomy, Uppsala University, Box
516, 751 20 Uppsala, Sweden}
\affiliation[15]{Instituto de Astrof\'isica e Ci\^encias do Espa\c{c}o, Universidade do Porto CAUP, 4150-762 Porto, Portugal}
\affiliation[16]{Departamento de F\'isica e Astronomia, Faculdade de Ci\^{e}ncias, Universidade do Porto, Rua do Campo Alegre 687, 4169-007 Porto, Portugal}
\affiliation[17]{Instituto de Física Teórica UAM-CSIC, Campus de Cantoblanco, 28049 Madrid, Spain}
\affiliation[18]{Institut de Recherche en Astrophysique et Plan\'etologie (IRAP), Universit\'e de Toulouse, CNRS, UPS, CNES, 14 Av. Edouard Belin, 31400 Toulouse, France}
\affiliation[19]{Universit\'e St Joseph; Faculty of Sciences, Beirut, BP-11514, Lebanon}

\emailAdd{debanjan.sarkar@mcgill.ca}
\emailAdd{mid.suman@gmail.com}

\abstract{
Line intensity mapping (LIM) has emerged as a powerful tool for surveying the large-scale structure of the Universe across cosmic time by measuring spatial fluctuations in the cumulative emission of spectral lines from unresolved sources or the intergalactic medium. Besides the most abundant 21-cm hyperfine line of neutral hydrogen, there are bright far-infrared fine-structure lines like [C\,\textsc{ii}] 158~$\mu$m, [O\,\textsc{iii}] 88~$\mu$m, [N\,\textsc{ii}] 122/205~$\mu$m, and [O\,\textsc{i}] 63~$\mu$m, as well as mid-/high-$J$ CO rotational transitions, hydrogen Ly$\alpha$ and H$\alpha$ as potential LIM probes. A key opportunity lies in combining and cross-correlating 21-cm intensity maps from SKAO with other line intensity maps, targeted by a range of ongoing and forthcoming LIM experiments that probe overlapping cosmic volumes. Cross-correlation between 21-cm maps and other line tracers mitigates uncorrelated systematics and enhances sensitivity to the underlying matter distribution, while multi-line analyses help disentangle cosmological and astrophysical parameters. Beyond cross-power spectra, higher-order and morphological statistics---such as cross-bispectra, marked correlations, and morphological measures---capture non-Gaussian features and the environmental dependence of structure formation. This chapter explores the synergies that can be achieved by combining SKAO observations with other line-intensity mapping experiments, demonstrating how such joint analyses can unlock new insights into galaxy evolution and cosmology.
}

\begin{document}

\include{journal-names}
\maketitle
\tableofcontents

\section{Introduction}
\label{sec:intro}

Line intensity mapping (LIM,~\cite{2017Kovetz,2022Bernal}) has emerged as a compelling technique to chart the three-dimensional large-scale structure of the Universe across a wide range of redshifts. By measuring the aggregate emission from spectral lines in unresolved galaxies, LIM bypasses the need to detect each galaxy individually. Instead, one observes intensity fluctuations on the sky and along the frequency dimension, which trace the underlying matter distribution. This approach allows extremely fast surveys over enormous cosmological volumes, capturing contributions from faint, numerous sources that would be inaccessible to pointed observations of individual galaxies. As such, LIM provides an unbiased probe of the cumulative emission and is particularly powerful for studying epochs and environments where only the brightest objects have been detected by traditional means. 

Among the various lines proposed for intensity mapping, the 21-cm spin-flip line of neutral hydrogen (\HI) stands out for its cosmological reach and scientific impact \citep{Madau:1996cs, Bharadwaj:2000av, Furlanetto:2006jb}. In the post-EoR (Epoch of Reionization) era ($z \lesssim 6$), most remaining \HI\ resides in damped Lyman-$\alpha$ (Ly$\alpha$) systems and galactic disks, making the 21-cm intensity signal an effective tracer of the distribution of galaxies and gas in and around halos~\citep{Bharadwaj:2000av, Bull:2014rha, Villaescusa-Navarro:2018vsg}. Fluctuations in the 21-cm brightness temperature directly reflect the underlying matter fluctuations on large scales, modulated by a bias factor and the mean \HI\ fraction \citep{Bagla:2009jy, Sarkar:2016lvb}. This enables a wealth of cosmological measurements, including the matter power spectrum shape, redshift-space distortions (RSD), and the imprint of baryon acoustic oscillations (BAO) that serve as a standard ruler for cosmic distances
\citep{Bharadwaj:2008yn, Sarkar:2018gcb, Sarkar:2019nak}. Crucially, 21-cm intensity mapping can probe very large scales ($\sim1000\,\mathrm{Mpc}$) and high redshifts, making them sensitive to effects like relativistic distortions \citep{Jolicoeur:2020eup} and primordial non-Gaussianity \citep{joudaki2011} in ways that optical galaxy surveys cannot.

The Square Kilometre Array Observatory (SKAO), with its unprecedented sensitivity and survey speed, will be a game-changer for 21-cm intensity mapping. SKA-Mid, with its dense array of mid-frequency radio dishes, is poised to deliver high-fidelity 21-cm maps over unprecedented areas and redshift ranges. In its Band~1 (approximately 350--1050~MHz), SKA-Mid will cover redshifts $z \sim 0.35$ up to $z \sim 3$, thus spanning almost the entire post-EoR epoch. This range captures the growth of structure from just after reionization through the peak of cosmic star formation ($z \sim 2$) down to the relatively mature Universe at $z \sim 0.3$.

Beyond 21-cm, a number of other spectral lines are being targeted for LIM, each illuminating different facets of galaxy evolution. The brightest line in the far-infrared is the [C\,\textsc{ii}] fine-structure line at 157.7~$\mu$m, which is emitted by singly ionized carbon in the cool (10--100~K) gas of photodissociation regions and also in diffuse neutral and ionized medium~\citep{Stacey:2010ps}. [C\,\textsc{ii}] is typically the brightest line in star-forming galaxies, containing up to $\sim 0.1\%$ of the total far-infrared luminosity, and hence is a premier target for high-redshift LIM. Similarly, [O\,\textsc{iii}] 88~$\mu$m, [O\,\textsc{i}] 63~$\mu$m, and [N\,\textsc{ii}] 122/205~$\mu$m lines probe H\,\textsc{ii} regions and other phases of the interstellar medium (ISM) associated with massive stars~\citep{Kewley_2019}. On the other hand, rotational transitions of carbon monoxide (CO) trace the cold molecular gas in galaxies~\citep{Kennicutt:2012ea}; the low-$J$ transitions such as CO(1--0), (2--1), (3--2) are of particular interest for $z \lesssim 6$ intensity mapping because they remain relatively bright and can be observed from ground-based facilities in the millimeter to centimeter bands (e.g. CO(1--0) at 30~GHz corresponds to $z\approx3.8$, CO(2--1) at 30~GHz corresponds to $z\approx7.6$, etc.). Moreover, hydrogen Ly$\alpha$ (121.6~nm rest,~\cite{Ouchi:2020zce}) and H$\alpha$ (656.3~nm rest) lines, while susceptible to absorption and scattering, trace star formation and the ionized intergalactic medium (IGM)~\citep{Kennicutt:1998zb}.

A key scientific opportunity in the LIM landscape lies in cross-correlating 21-cm intensity maps from SKAO with other LIM datasets probing the same cosmic volumes. Even when auto-spectra of individual lines are compromised by foreground contamination or instrumental systematics, cross-correlations can robustly extract the shared cosmological signal~\citep{Lidz:2011dx, Furlanetto:2006pg, McBride:2023exl}. This technique inherently suppresses uncorrelated noise and foregrounds -- such as Galactic synchrotron in 21-cm or Galactic dust in far-infrared lines -- and has already been used to achieve the first statistical detections of the 21-cm signal in MeerKAT~\citep{Booth:2009ex} at $z \sim 0.8$ via cross-correlation with the WiggleZ galaxy survey~\citep{Cunnington:2022uzo}. By measuring cross-power spectra between multiple line pairs \citep{Schaan:2021gzb}, one can also form ratios that cancel out nuisance astrophysical parameters.
Moreover, going beyond two-point functions, higher-order statistics \citep{Majumdar01.2026.SKA} such as the cross-bispectrum \citep{Moodley:2023lmu} and marked correlations \citep{Kamran:2024xob} can capture non-Gaussian features and environmental dependencies, offering deeper insight into galaxy evolution and structure formation. Further, image-domain morphological measures, such as local dimension \citep{Sarkar:2008bm, dosibhatla_2025_lss-morphology} and largest cluster statistics \citep{klypin_1993_percolation, Dasgupta:2023fyf, dosibhatla_2025_lss-morphology}, can access non-Gaussian and morphological information.
These approaches are central to maximizing the scientific return of LIM surveys in the SKAO era.

In this chapter, we review the scientific potential of multi-wavelength LIM synergies anchored by SKAO. We begin by surveying LIM target lines beyond the 21-cm transition and outline the current and upcoming experiments that are particularly relevant for synergy with SKAO.  We define and discuss different statistical estimators necessary for interpreting the joint datasets, focusing on cross-power spectra, cross-bispectra, null tests such as the $\mathcal{Q}$ estimator, and marked correlation functions. 
We further summarize the potential of these surveys in jointly constraining astrophysics and cosmology by estimating the corresponding signal-to-noise ratios.
Finally, we offer a forward-looking perspective on the evolving LIM landscape. We assess the challenges that must be overcome, and highlight promising strategies to mitigate them, emphasizing approaches that preserve cosmological information under realistic observational conditions. SKAO will be central to unlocking the full promise of LIM as a probe of cosmic structure and galaxy evolution.

\section{LIM with spectral lines beyond 21-cm in the post-EoR Universe}
\label{sec:other_lines}

Lines captured by LIM experiments encode complementary information. In this Section, we summarize their key properties, and we summarize which surveys are dedicated to their observations.

\subsection{Key target lines and their tracers of galaxy evolution}
A variety of spectral lines from the ISM complement the 21-cm line by tracing different gas phases:

$\bullet$ \textbf{[C\,\textsc{ii}] 158~$\mu$m:} This line originates from the fine structure $2P^{3/2}$ to $2P^{1/2}$ transition from singly ionized carbon, which is abundant in photodissociation regions (PDRs) \citep{Hollenbach_1999_PDR} at the interface of H\,\textsc{ii} regions and molecular clouds. [C\,\textsc{ii}] can also originate from diffuse cold neutral medium and the warm ionized ISM, and even from CO-dark molecular gas. Because it is one of the strongest cooling lines in star-forming galaxies, [C\,\textsc{ii}] broadly traces star formation activity and correlates with the star formation rate (though with scatter and systematic variation due to metallicity and ISM properties). At high-redshifts ($z \sim 4$–6), [C\,\textsc{ii}] has been detected in individual galaxies using ALMA (e.g. the ALPINE survey, \citealt{Schaerer+2020}), and LIM experiments aim to measure its cumulative intensity \citep{Lagache2018,Roy:2024dmt, Murmu_2021, Murmu_2023}.

$\bullet$ \textbf{CO rotational lines:} Carbon monoxide is the classic tracer of molecular hydrogen ($\mathrm{H}_2$) in galaxies. The lowest transition, CO(1--0) at rest 115.3~GHz, is directly proportional to molecular gas mass through a conversion factor (which may vary with environment). Higher-$J$ transitions (CO(2--1), CO(3--2), etc.) trace progressively warmer and denser molecular gas; in starburst environments these higher lines can be significantly excited. For $z \lesssim 6$, CO(1--0) and CO(2--1) are prime LIM targets because of their relative brightness and accessibility in the cm-/mm-wave atmospheric windows. Mapping the CO intensity unveils the total molecular gas distribution and its evolution, which is essential for understanding the gas supply for star formation over cosmic time. The single-dish LIM experiment COMAP~\citep{Keating:2020wlx}, currently in its Pathfinder phase (see details in Sec. 2.3), recently published the first direct limits on the CO(1-0) power spectrum at $z = 2.4-3.4$~\citep{COMAP:2022,Stutzer:2024rps}.

$\bullet$ \textbf{[O\,\textsc{iii}] 88~$\mu$m and other FIR lines:} Doubly ionized oxygen emits [O\,\textsc{iii}] 88~$\mu$m in H\,\textsc{ii} regions around young massive stars. This line tends to be strong in low-metallicity galaxies and is a useful probe of the ionized gas conditions. Meanwhile, neutral oxygen’s [O\,\textsc{i}] 63~$\mu$m arises from dense PDRs, and [N\,\textsc{ii}] 122~$\mu$m and 205~$\mu$m come from ionized regions (with [N\,\textsc{ii}] 205~$\mu$m primarily from diffuse ionized gas). While individually fainter than [C\,\textsc{ii}], these lines provide additional constraints on ISM phases (e.g. [N\,\textsc{ii}] traces H\,\textsc{ii} region volume and thus the density of ionized gas).

$\bullet$ \textbf{Ly$\alpha$ 121.6~nm and H$\alpha$ 656.3~nm:} These are hydrogen recombination lines. Ly$\alpha$ is a tracer of the global ionization state and a difficult line to interpret due to radiative transfer effects. H$\alpha$, by contrast, is a non-resonant recombination line that directly traces on-going star formation (through ionizing photon production). Intensity maps of Ly$\alpha$ and H$\alpha$ can reveal the global distribution of ionizing sources and the ionized IGM, especially during and after reionization. Recently, the Hobby-Eberly Telescope Dark Energy Experiment  (HETDEX~\citep{Gebhardt:2021vfo}) probed the cross correlation between Ly$\alpha$ intensity fluctuations and Ly$\alpha$ emitting galaxies at $z=1.9-3.5$~\citep{Niemeyer:2025yvq}.

\subsection{Experiments targeting [C\,\textsc{ii}] 158~$\mu$m intensity}
The EXperiment for Cryogenic Large-Aperture Intensity Mapping (EXCLAIM) is a pathfinder mission focusing on the [C\,\textsc{ii}] 158~$\mu$m line in the post-reionization Universe. EXCLAIM is a balloon-borne telescope equipped with a Fourier transform spectrometer covering frequencies 420--540~GHz at a spectral resolution of $R \approx 512$ \citep{Pullen+2023}. This band targets redshifted [C\,\textsc{ii}] in the range $z \sim 2.5$ to 3.5. The choice of this redshift window corresponds to a time just before the peak of cosmic star formation ($z \sim 2$) when the [C\,\textsc{ii}] intensity is expected to be substantial. EXCLAIM’s survey plan covers about $305~\mathrm{deg}^2$ along the celestial equator (overlapping the Sloan Digital Sky Survey Stripe 82 field). With this broad area, EXCLAIM aims to measure the [C\,\textsc{ii}] intensity power spectrum and possibly cross-correlations with external datasets. Indeed, the EXCLAIM fields have planned overlap with galaxy redshift surveys (e.g. BOSS quasars and emission-line galaxies, HETDEX Ly$\alpha$ emitters, and deep optical imaging from HSC), enabling cross-correlation of [C\,\textsc{ii}] intensity maps with known galaxies. If SKA-Mid conducts single-dish 21-cm intensity mapping in a portion of the same region (and at redshifts overlapping $z\sim2.5$–$3.5$, corresponding to $\nu \sim 350$–$450$~MHz), there is an enticing possibility of a 21-cm–[C\,\textsc{ii}] cross-correlation. Such a measurement would directly connect the neutral atomic gas with the star-forming ISM in the same volumes, providing a proof-of-concept for multi-line studies. EXCLAIM’s sensitivity will be modest (as a short-duration balloon experiment), but it is an important pilot for later satellite or ground-based [C\,\textsc{ii}] intensity mapping efforts.
\par
The Fred Young Submillimeter Telescope (FYST)~(\cite{CCAT_2023}) is another experiment focused on CII intensity mapping. FYST plans to conduct a deep survey, covering two fields of approximately $4~\mathrm{deg}^2$ each, with a total of 2000 hours of observation. The telescope will operate at frequencies ranging from roughly 210 to 430 GHz, with a primary emphasis on CII line intensity mapping at redshift values between 3.5 and 8.05. This setup enables the cross-correlation of the 21-cm signal from SKA-Low with the CII intensity maps at the end of the reionization.
\subsection{Experiments targeting CO intensity at the epoch of galaxy assembly}
The CO Mapping Array Project (COMAP) is a dedicated program to map CO line intensity fluctuations. The current COMAP Pathfinder instrument operates in the 26-34~GHz band, targeting CO(1--0) emission from galaxies at $z \sim 2.4$-3.4 \citep{Cleary+2022}. This epoch corresponds to the so-called “Cosmic Noon” when the star formation rate density of the Universe was near its peak and molecular gas was abundant. The COMAP Pathfinder consists of a 10.4-meter dish outfitted with a multi-beam cryogenic receiver, and it has been surveying several fields totaling a few square degrees (one of which lies in the HETDEX Spring field in the northern sky). The first and second seasons of COMAP observations have placed upper limits on the CO power spectrum at $z \sim 3$ \citep{Ihle_2022_COMAP_PS, Stutzer:2024rps}, already constraining the product of CO luminosity and bias of emitting galaxies \citep{Chung_2022_COMAP_constraints, Chung_2024_COMAP_constraints}. 
COMAP aims to detect the CO clustering signal and measure both the clustering component and the shot noise (sensitive to the abundance of very luminous CO emitters); additional seasons are already funded. 
Beyond the Pathfinder, COMAP has plans for a Stage-II expansion, to widen the survey area and include higher-frequency receivers to target CO(2–1) or other lines, and eventually a next-generation experiment aiming even at CO during reionization-era ($z\sim 7$) dubbed “COMAP-EoR”~\citep{COMAP:2021nrp}. 

These efforts align well with SKAO’s timeline, meaning that by the time SKA-Mid is carrying out a Band 1 21-cm survey, COMAP and its successors could provide ample opportunities for cross-correlation. Although there may not be a direct overlap between the planned COMAP survey fields and the fields to be surveyed by SKAO at the moment, one can always envisage a dedicated survey plan or the construction of a COMAP-like experiment with similar instrumental performance and survey parameters that have significant overlap with the surveys from SKAO. Such a cross-correlation would be valuable for both confirming the presence of both signals (since a cross-detection could be achieved even if neither auto-spectrum is independently significant) and for studying the relationship between atomic and molecular gas in galaxies. Additionally, because CO and HI biases may differ, the cross-spectrum would help break degeneracies in each dataset; e.g.,~a joint analysis could separate the mean CO luminosity density and bias from the combination of auto and cross spectra, whereas each alone leaves a degeneracy between luminosity and bias.

The Tomographic Ionised-carbon Mapping Experiment (TIME) \citep{Sun_2021_TIME} is a spectrometer array designed to be mounted on an ALMA $12$m Prototype Antenna at the Arizona Radio Observatory (ARO). The instrument operates in a frequency range of approximately $200-300$ GHz, and its primary target line is the CII line at $5.29 < z < 8.51$. However, TIME can also map several CO lines at lower redshifts, viz. the CO(3--2) line at $0.15 < z < 0.73$, CO(4--3) line at $0.53 < z < 1.31$, CO(5--4) line at $0.91 < z < 1.88$, etc. The CO intensity mapping signal from TIME falls within overlapping redshifts with the 21-cm signal from SKA-Mid and can be employed for cross-correlation. However, the strong C \textsc{ii} signal from the end of reionization must be accounted for accurately to isolate the CO emissions.

\subsection{Space-based LIM experiments targeting Ly$\alpha$, H$\alpha$, and related lines}
Complementing the ground-based sub-mm and radio experiments, the SPHEREx (Spectro - Photometer for the History of the Universe, Epoch of Reionization, and Ices Explorer) SPHEREx satellite~\citep{SPHEREx:2014bgr}, launched in March 2025, will provide all-sky intensity maps in multiple optical to near-infrared bands. SPHEREx will conduct an all-sky survey with spectral resolution $R \sim 40$ over wavelengths $0.75$--5~$\mu$m \citep{Crill+2020}. Although its primary science goals include studies of inflation and Galactic ices, SPHEREx will deliver a rich dataset for cosmology and galaxy evolution, including intensity maps of lines such as H$\alpha$ (traced by a blend of narrow bands) out to $z \approx 5$, and the Pa$\alpha$/Pa$\beta$ lines at lower $z$, among others. The line of most interest for high-$z$ LIM is likely Ly$\alpha$ at $z \sim 6$–$10$, whose redshifted wavelength (0.75–1~$\mu$m) falls in SPHEREx’s range. However, detecting Ly$\alpha$ fluctuations is challenging due to foregrounds and interloper lines (especially the ubiquitous O\,\textsc{ii} 372.7~nm line from lower redshifts). Still, SPHEREx data can be used statistically to extract line intensity information, and cross-correlation with 21-cm or [C\,\textsc{ii}] could help isolate Ly$\alpha$ (since 21-cm and [C\,\textsc{ii}] at high $z$ would not correlate with low-$z$ O\,\textsc{ii} emission). Moreover, at lower redshifts ($z<2$), SPHEREx will map H$\alpha$ intensity, which can be cross-correlated with 21-cm intensity maps \citep{Fonseca_2018} in the same volume (H$\alpha$ at $z\sim1$ appears around 1.3~$\mu$m, within SPHEREx, and the same volume’s 21-cm signal would be at $\nu \sim 710$~MHz, accessible to SKAO Band~2 or Band~1). Such a cross-correlation could constrain the bias and abundance of H$\alpha$ emitters and serve as an independent check of the 21-cm auto-spectrum.

Another concept, the Cosmic Dawn Intensity Mapper (CDIM,~\cite{cooray_2019}), has been proposed to map Ly$\alpha$, H$\alpha$, and other lines at higher spectral resolution ($R \sim 300$) over a wide field, specifically targeting the reionization era and cosmic dawn. If realized, CDIM or similar experiments would be ideal counterparts to SKAO 21-cm surveys of reionization (with SKA-Low at $z>6$) and to SKA-Mid at lower redshifts.
Further information on the $z\sim 1-2$ Ly$\alpha$ emission will be provided by the upcoming Ultraviolet Transient Astronomy Satellite (ULTRASAT,~\cite{Shvartzvald:2023ofi}) (launch planned in 2027). Although designed to observe transients, ULTRASAT will also provide a near-UV (230-290 nm) full-sky map that will be analyzed with LIM tools~\citep{Libanore:2024wvv}, thus providing other means for cross-correlation studies.

\subsection{Survey overlap strategy for multi-line synergies}
\label{sec:overlap_strategy}
\begin{figure}[htbp]
    \centering
    \begin{subfigure}[t]{0.64\textwidth}
        \centering
        \includegraphics[width=\textwidth]{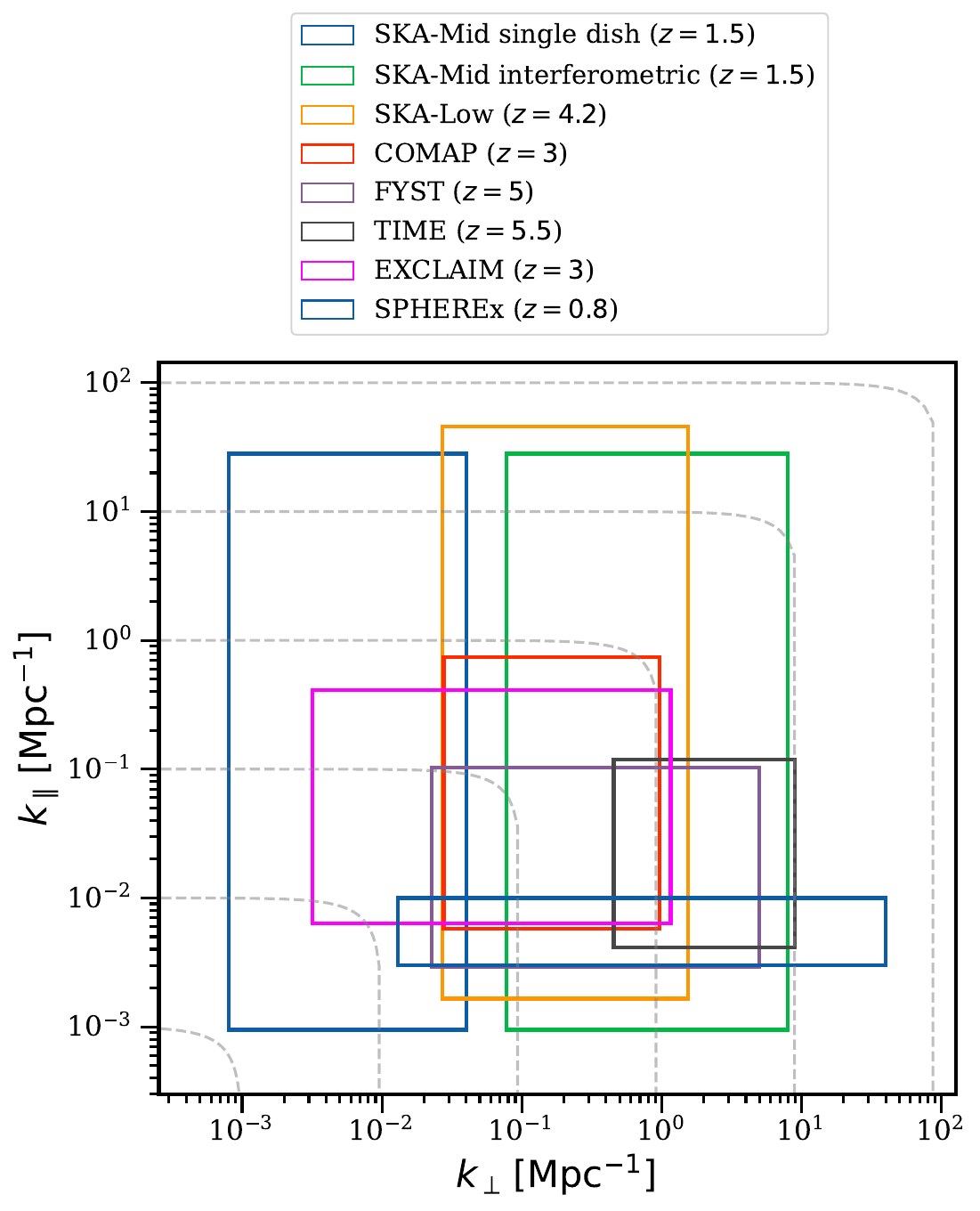}
    \end{subfigure}
    \hfill
    \begin{subfigure}[t]{0.35\textwidth}
        \centering
        \includegraphics[width=\textwidth]{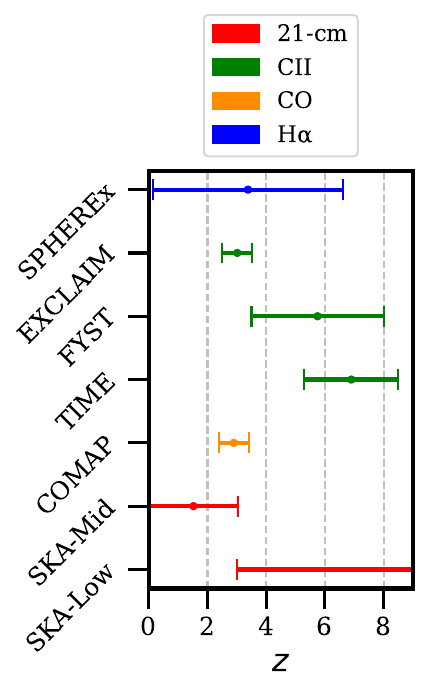}
        
    \end{subfigure}
    \caption{\textbf{Left:} The longitudinal ($k_\parallel$) and transverse ($k_\perp$) Fourier mode coverage of various LIM surveys. The numbers are approximate, and the $k_\perp$ values are calculated for a fixed redshift mentioned alongside the survey name. Spherically averaged $k$-values are indicated by dashed grey curves. The actual coverage on the $k$-plane may vary depending on the cuts applied to frequency channels, baselines or survey area. \textbf{Right:} The redshift coverage of different LIM surveys. The actual redshift values may vary depending on frequency cuts.}
    \label{fig:coverage_plots}
\end{figure}

Realizing the full potential of multi-wavelength intensity mapping requires coordinating surveys so that they overlap on the sky and in redshift. Only with sufficient volumetric overlap can cross-correlations be measured with high significance. A practical strategy is as follows. First, identify target redshift ranges that are scientifically rich and where multiple tracers are accessible. The right panel of Figure \ref{fig:coverage_plots} shows the ranges of redshift coverage of different LIM surveys. One promising redshift range is $z \sim 2$–3: here 21-cm is observable in SKA-Mid Band~1; [C\,\textsc{ii}] can be mapped by experiments like  EXCLAIM  or a space mission; CO(1--0) can be observed by a COMAP-Like experiment and there exist extensive galaxy surveys (optical/infrared) that can be used for cross-checks or additional cross-correlations. Another attractive range is $z \sim 5$–6, around the end of reionization, where 21-cm (from SKA-Low) could be cross-correlated with [C\,\textsc{ii}] intensity maps (from experiments like CONCERTO or TIME) and with galaxy surveys (JWST or ground-based high-$z$ galaxy searches). In each case, careful matching of survey footprints is needed. Ideally, one would have a contiguous field (or a set of fields) that all instruments observe. This has been the approach in the HETDEX field, for example, which many LIM and galaxy surveys target. Shared fields like COSMOS or Stripe~82 can also serve as coordination points.

In terms of observing strategy, large-area shallow surveys are generally preferable for cross-correlation over small deep fields, because the largest angular scales provide the highest signal-to-noise for correlated fluctuations (and cosmic variance on large modes is reduced when combining two tracers compared to one). However, the choice also depends on signal amplitude and experiment noise: for a given total observing time, there may be an optimal balance between area and depth. These trade-offs are being studied by each project, but from the standpoint of cross-correlation, it is important that neither survey is dramatically smaller than the other. The left panel of Figure \ref{fig:coverage_plots} shows the Fourier mode coverage of different LIM surveys. If one survey covers a tiny field embedded in a much larger field of the other, cross-correlation modes are limited to the smaller area. Thus, coordination in survey design, at least reserving some overlapping area, is crucial.

Another consideration is spectral resolution and redshift binning. Cross-correlating intensity maps requires overlapping redshift windows typically of width $\Delta z/(1+z) \lesssim$ a few percent (to ensure the structures truly overlap along the line of sight). If one survey has coarse redshift resolution (e.g. a broad tomographic bin) and the other has fine resolution, one need to degrade the latter to match the former. Ideally, experiments targeting cross-correlation should have comparable spectral resolution. 
In practice, one can convolve or rebin one dataset to match the other’s resolution for the cross-correlation analysis.

In this chapter, we will highlight a few benchmark scenarios of overlapping surveys and estimate the achievable signal-to-noise for cross-correlations in those cases. These examples will illustrate how the design of overlapping surveys (areas, depths, and resolutions) translates into detection significance for the cross-spectra and what science can be extracted.

\section{Joint constraints on the cosmology and astrophysics via different LIM observations and statistics}
\label{sec:joint_cons}

A central advantage of LIM is its capacity to deliver joint constraints on cosmology and astrophysics, which is crucial given the inherent degeneracies between these two domains in observational data \citep{2021Schaan, 2019Bernal}.
LIM data are highly valuable for calibrating a wide array of astrophysical processes and parameters, such as star formation rates and the evolutionary pathways of galaxies~\citep{2021Schaan, 2019Bernal}, particularly at high redshifts \citep{2022Bernal, 2023Sato}. At the same time, 
LIM is capable of refining fundamental cosmological parameters, including the sum of neutrino masses~\citep{MoradinezhadDizgah:2021upg,Shmueli:2024npx}, and the amplitude of primordial non-Gaussianity~\citep{2021Schaan, 2019Bernal}. It can also provide information on the dark energy equation of state, often symbolized as $w$. This is a critical parameter for understanding the accelerating expansion of the universe, and joint cosmological constraints on dark energy models are typically derived from combining various observational datasets, such as measurements of the Hubble parameter, Type Ia supernovae (SNIa), Baryon Acoustic Oscillations (BAO), and the Cosmic Microwave Background (CMB)~\citep{2021Schaan}. 
Furthermore, LIM has been proved to be indirectly sensitive to the abundance of low mass halos; forecasts show that LIM will provide unprecedented constraints on the amplitude of the power spectrum on small scales ($\sim 10$\,Mpc)~\citep{Libanore:2022ntl}, hence bringing invaluable information to constrain beyond-$\Lambda$CDM cosmological models, e.g.,~fuzzy dark matter or primordial magnetic fields~\citep{Adi:2023qdf}. Wide sky LIM surveys can also be sensitive to signatures of exotic particles, such as decaying dark matter~\citep{Bernal:2020lkd,Libanore:2024hmq}.

Joint statistical analyses, particularly those involving multi-line observations and cross-correlations with other surveys, are essential for breaking degeneracies between astrophysical and cosmological parameters, leading to significantly tighter constraints than individual surveys alone. For example, cross-correlating Carbon Monoxide (CO) LIM signals with spectroscopic and photometric galaxy surveys has been shown to yield detectable signals, thereby enhancing the understanding of both large-scale structure formation and small-scale astrophysical phenomena~\citep{2019Chung}. The multi-tracer intensity mapping technique, which involves cross-correlations between different line intensity maps or between intensity maps and galaxy/CMB lensing, is an emerging method that facilitates the construction of consistent halo models for joint analyses in both 3D redshift-space and 2D projected maps~\citep{2023Sato}. These approaches improve the precision of parameter inference beyond what is achievable with traditional methods. Moreover, to mitigate the degeneracies between astrophysical and cosmological parameters, sophisticated modeling frameworks can be employed, incorporating phenomena such as redshift-space distortions (RSD) and the Alcock-Paczynski (AP) effect in the modeling of the power spectrum. RSD, caused by the peculiar velocities of galaxies, modify the observed clustering patterns in redshift space and are instrumental in measuring the growth rate of large-scale structure. Sufficient spectral resolution in LIM observations enables the measurement of the large-scale RSD in the two-halo term of the power spectrum, thereby allowing for simultaneous constraints on cosmological parameters. 

SKAO will foster our understanding of cosmology and astrophysics, both alone and in cross correlation studies. Surveys in its Band~1 will constrain how the neutral hydrogen mass is distributed across halos over time, and will yield precise measurements of cosmological parameters governing the expansion and growth of structure (e.g. the growth rate $f\sigma_8$, the Hubble parameter $H(z)$ via BAO, and potential deviations from $\Lambda$CDM). 
Forecasts indicate that SKAO 21-cm LIM, in both single-dish and interferometric modes, will yield high-precision measurements of the power spectrum and its multipoles, providing constraints on cosmological parameters that rival or surpass current probes such as the CMB and galaxy surveys \citep{2023Berti, 2019Zhang}. Simulated results further show that combining SKAO 21-cm LIM data with CMB observations (e.g., Planck) could reduce uncertainties on key parameters like the Hubble constant $H_0$ and dark matter density $\Omega_c h^2$ by up to a factor of four compared to Planck alone. Moreover, SKAO observations are expected to break existing parameter degeneracies, particularly between matter density and the Hubble constant, thereby tightening constraints on dark energy parameters and opening new avenues for probing non-standard physics, including neutrino masses and dark energy evolution \citep{2019Zhang, 2023Berti}.


\section{Cross-correlation statistics and associated science}

\subsection{Cross-power spectrum}
\label{sec:cross-corr}
Given multi-wavelength LIM observations of the Universe at the same redshifts, it is possible to cross-correlate the observed LIM maps and estimate a statistic known as the cross-power spectrum. This is defined as,
\begin{equation}
    \big\langle \delta \tilde{I}_{\rm a}(\bm{k}) \delta \tilde{I}^*_{\rm b}(\bm{k^\prime})\big\rangle = V \delta_{k,k^\prime} P_{\rm a\times b}(k),
\end{equation}
where $\delta \tilde{I}_{\rm a}(\bm{k})$ and $\delta \tilde{I}_{\rm b}(\bm{k})$ are the fluctuations for LIM signals, a and b, in the Fourier domain. $V$ is the cosmological volume over which the LIM signals are observed, and $P_{\rm a\times b}(k)$ is the cross-power spectrum between these two LIM signals. The unique aspect of the cross-power spectrum over the auto-power spectrum is that it retains the relative phase difference between the LIM signals. If the fluctuations of the LIM signals, a and b, in the Fourier domain are written as $\delta\tilde{I}_{\rm a} = a\exp(-\bm{i}\phi)$ and $\delta\tilde{I}_{\rm b} = b\exp(-\bm{i}\phi)$, then one can see that $\big\langle \delta\tilde{I}_{\rm a}\delta\tilde{I}_{\rm b}^*\big\rangle = \langle a b\exp({-\bm{i}(\phi_{\rm a} - \phi_{\rm b})})\rangle$, where $\phi_{\rm a}$ and $\phi_{\rm b}$ are the phases of the LIM signals in the Fourier domain. Therefore, to quantify the strength of correlation between the LIM signals, one can define the cross-correlation coefficient as
\begin{equation}
    r_{\rm a\times b}(k) = \frac{P_{\rm a\times b}(k)}{\sqrt{P_{\rm a}(k)P_{\rm b}(k)}}
\end{equation}
where $P_{\rm a}(k)$ and $P_{\rm b}(k)$
are the power spectra of the respective LIM signals.\par
Earlier works, such as~\citet{Silva+2013}, have explored the prospects for cross-correlating LIM signals for Ly-$\alpha$ emission from galaxies and the 21-cm emission from the IGM during the EoR regime. Since these two LIM signals probe different types of matter distribution (galaxies and IGM) in the Universe, their cross-power spectrum, as given by~\cite{Silva+2013}, can be written as
\begin{equation}
    P_{\rm Ly\alpha\times 21-cm}(k) = \langle I_{\rm Ly\alpha - {\rm GAL}}(z) \rangle\langle I_{\rm 21-cm-{\rm IGM}}(z) \rangle\bigg[P_{\delta\delta}(k,z) - \frac{1}{1 - \langle x_i(z)\rangle}P_{x_i\delta}(k,z)\bigg].
\end{equation}
Here, $\langle x_{i}(z)\rangle$ is the average ionized fraction and $P_{x_i\delta}(k,z)$ is the cross-power spectrum of the ionized fraction field and the matter field.

Recently, \cite{Libanore:2025wtu} introduced an alternative formalism to correlate multiple lines and produce their intensity map fluctuations during the EoR and at lower $z$, extending the analytical treatment developed in ~\cite{Munoz:2023kkg} for the 21-cm auto-correlation.

\par
In the post-EoR regime, most of the IGM is ionized. Therefore, the line emissions originating from this regime are expected to trace the galaxies, which are the primary sources of various line emissions. Therefore, cross-correlating two different LIM signals of line emissions originating from galaxies in the post-EoR regime is expected to probe the distribution of galaxies and also the mutual correlation between individual line emissions. If we consider two line emissions, $i$ and $j$, originating from the galaxies, the cross-power spectrum can be written as
\begin{equation}
    P_{i\times j}(k,\mu,z) = P^{2{\text -}\rm halo}_{i\times j} (k,\mu,z) + P^{1{\text -}\rm halo}_{i\times j} (k,\mu,z) + P^{\rm shot}_{i\times j}(z)
\end{equation}
~\citep{Schaan+2021}.
The first term, known as the two-halo term, is given as
\begin{equation}\label{eq:P2halo}
    P^{2{\text -}\rm halo}_{i\times j} (k,\mu,z) = \langle I_{i}(z)\rangle\langle I_{j}(z)\rangle\big[b_{i}(k,\mu,z) + F\mu^2\big]\big[b_{j}(k,\mu,z) + F\mu^2\big]P_{\rm m}(k,z).
\end{equation}
Here, $\langle I_{i}(z)\rangle$ is the mean line intensity of the $i$-th line originating from the galaxies, given as
\begin{equation}
    \langle I_{i}(z)\rangle = \frac{1}{4\pi\nu^0_{i}}\frac{c}{H(z)}\int dL_{i}\,\Phi(L_{i}, z)L_{i};
\end{equation}
$\nu_i^0$ is the rest-frame frequency of the line emission, $L_i$ is the luminosity of the $i$-th line and $\Phi(L_i,z)$ its luminosity function. In Eq.~\eqref{eq:P2halo}, $b_i(k,z)$ is the line effective bias, defined by averaging the contribution of the line sourced in halos with different masses $M$. Labeling $dn(M)/dM$ as the halo mass function, $b(M)$ as the halo bias and $u(k,M)$ as the halo density profile, we can write
\begin{equation}
    b_{i}(k,\mu,z) = \bigg(\int dL_{i}\,\Phi(L_i,z)L_{i})\bigg)^{-1}\int dM \frac{dn(M)}{dM}L_i(M)b(M)u(k,M)\exp(-k^2\mu^2\sigma^2_d(M)/2).
\end{equation}
On the other hand, $F$ is the effective growth rate in the redshift space, given as
\begin{equation}
    F(k,\mu,z) = f \int dM \frac{dn(M)}{dM} \bigg(\frac{m}{\langle \rho\rangle}\bigg)u(k, M)\exp(-k^2\mu^2\sigma^2_d(M)/2),
\end{equation}
where $f \equiv d\ln D/d \ln a$, $\langle\rho\rangle$ the average density, and $\sigma_d$ is the pairwise velocity dispersion. $P_{\rm m}(k,z)$ is the power spectrum of the underlying matter distribution. The second term, which is the 1-halo term, is given as
\begin{equation}
    P^{1{\text -}\rm halo}_{i\times j}(k,\mu,z) = \bigg(\frac{c}{4\pi H(z)}\bigg)^2\frac{1}{\nu^0_i\nu^0_j}\int  dM \frac{dn(M)}{dM}|u(k, M)|^2\exp(-k^2\mu^2\sigma^2)L_i(M)L_j(M).
\end{equation}
It can be seen that the correlation between the lines $i$ and $j$ at the same halo mass contributes to this term. Finally, the shot noise term, $P^{\rm shot}_{i\times j}(k,z)$ can be written as
\begin{equation}
    P^{\rm shot}_{i\times j} = \bigg(\frac{c}{4\pi H(z)}\bigg)^2\frac{1}{\nu^0_i\nu^0_j}\int dL_idL_j\,\Phi(L_i,L_j)L_iL_j.
\end{equation}
$\Phi(L_i,L_j) = \int dM\,dn(M)/dM\,\phi(L_i,L_j|M)$ is the bivariate luminosity function with line luminosities $L_i$ and $L_j$, and $\phi(L_i,L_j|M)$ is the conditional luminosity function for halo mass $M$. Thus, for the shot noise term, the correlation information for lines $i$ and $j$ originating from the same galaxy is captured in the bivariate luminosity function.

\subsection{21-cm-galaxy cross-bispectrum}
The power spectrum measures the amplitude of the signal fluctuations at different scales and only completely characterizes the statistical properties of a signal if it is a Gaussian random field. Due to the gravitational instability, the redshifted 21-cm signal from the post-Reionization (post-EoR) will be highly non-Gaussian \citep{Sarkar:2019ojl}. Additionally,
the signal will contain the imprints of the primordial non-Gaussianity. Thus, one has to choose a statistic sensitive to the non-Gaussianity inherently present in the 21-cm signal. To achieve this, one has to consider higher-order statistics such as the bispectrum \citep{Majumdar01.2026.SKA}. The residual foreground and systematics associated with the instrument will hinder a high signal-to-noise ratio detection of the 21-cm auto bispectrum with fewer observational hours. One way to boost the signal-to-noise ratio is to do a cross-correlation measurement with other line intensity mapping experiments focused on \CII\;$\mu$m, \CO\ or galaxy redshift surveys. Additionally, the cross-correlation signal can put competitive constraints on cosmological parameters comparable to those of auto-correlation \citep{2024Berti}.
\par
The cross-bispectrum of the three fields $F_{1}$, $F_{2}$ and $F_{3}$ can be defined as 
\begin{align}
\langle \Delta_{F_1}(\mathbf{k_1}) \Delta_{F_2}(\mathbf{k_2}) \Delta_{F_3}(\mathbf{k_3}) \rangle = V \delta_{\mathbf{k_1} + \mathbf{k_2} + \mathbf{k_3},0}~B_{F_1,F_2,F_3}(\mathbf{k_1},\mathbf{k_2},\mathbf{k_3}),
\end{align}
where $\Delta_{F_1}$, $\Delta_{F_2}$ and $\Delta_{F_3}$ are the Fourier transform of the fields $F_1$, $F_2$ and $F_3$, respectively.
$\delta_{\mathbf{k_1} + \mathbf{k_2} + \mathbf{k_3},0}$ is the Kronecker delta function, and its numerical value is equal to one when the condition  $\mathbf{k_1} + \mathbf{k_2} + \mathbf{k_3}=0$ is satisfied and zero otherwise.
The cross-bispectrum of the signals $F_1$,$F_2$, and $F_3$ can be estimated for various combinations. If we only consider the cross-bispectrum of the two signals, this results in six cross-combinations and they are $B_{F_1, F_2, F_2}$, $B_{F_2, F_1, F_2}$, $B_{F_2, F_2, F_1}$, $B_{F_1, F_1, F_2}$, $B_{F_1, F_2, F_1}$, $B_{F_2, F_1, F_1}$.
Considering the cross-bispectrum of the 21-cm signal and galaxy clustering, one can have cross-bispectrum with cross-combination containing two HI fields and one galaxy field, and one HI field and two galaxy fields. The bispectrum can be estimated for different sizes and shapes of the $k$-triangles in Fourier space. To identify all the unique shapes of the triangles in Fourier space one can use the bispectrum parametrization introduced in \cite{Bharadwaj_2020} . According to this parametrization, for a triangle in Fourier space with $k_1\leq k_2 \leq k_3$, its size is determined by $k_1$ and the shape by $k_2/k_1$ and the cosine of the angle ($\cos \theta$) between $k_1$ and $k_2$. 
\par
Here, we show a forecast for the detectability of the 21-cm-galaxy cross-bispectrum for SKA-Mid and Euclid-like galaxy redshift survey \citep{Noble:2026oqm}. The 21-cm line intensity maps and Euclid-like mock galaxy catalogues for the forecast were generated by post-processing the galaxy catalogues from the GAEA semi-analytic galaxy formation model~(\citealt{DeLucia_2014, Hirschmann_2016, DeLucia_2024, Fontanot_2025}, and references therein). To estimate the 21-cm-galaxy cross-bispectrum, we used the fast Fourier transform (FFT) based estimator presented in \cite{Shaw_2021}, which parameterizes bispectra following the unique triangle shape classification of \cite{Bharadwaj_2020}. 
\begin{figure}[htp]
    \centering
    
    \begin{subfigure}{1\textwidth}
        \centering
        \includegraphics[width=\linewidth]{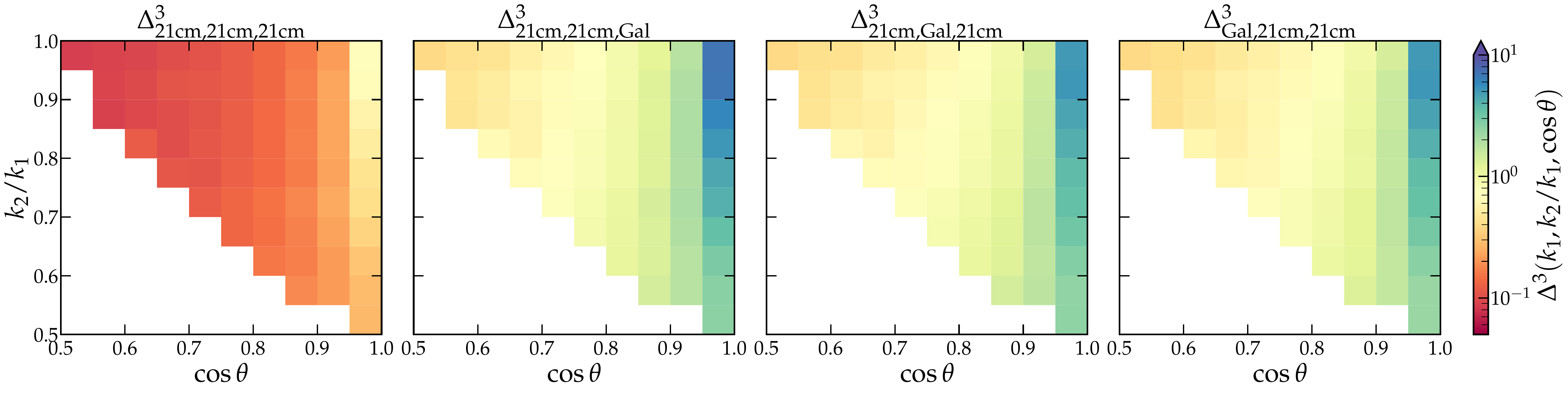}
        \label{fig:sub1}
    \end{subfigure}
    \begin{subfigure}{1\textwidth}
        \centering
        \includegraphics[width=\linewidth]{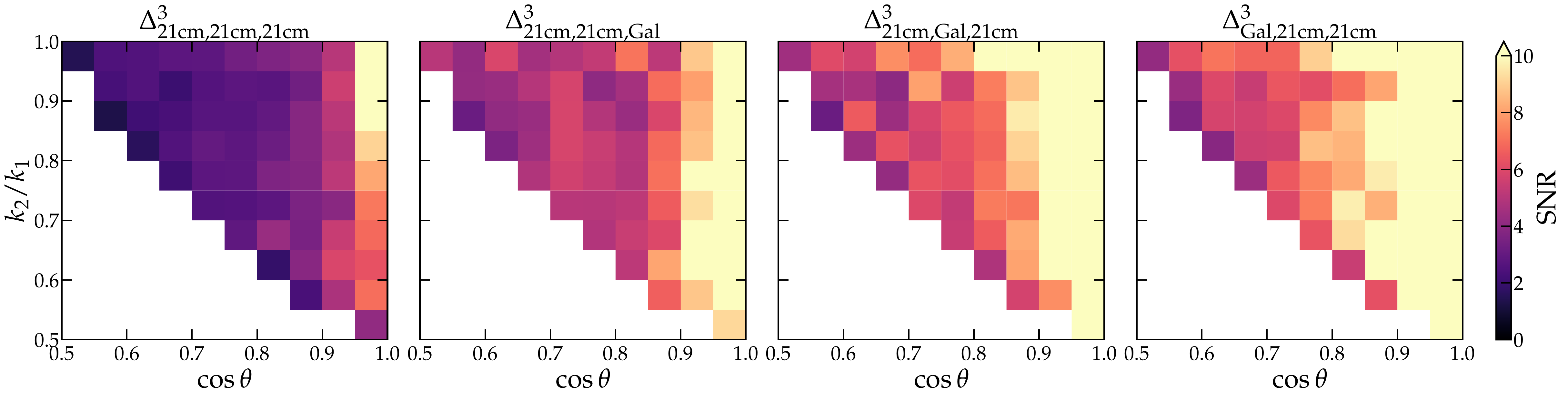}
        \label{fig:sub2}
    \end{subfigure}
    
    \caption{\textbf{Top panel:} The 21-cm auto-bispectrum and 21-cm-galaxy cross-bispectrum for all unique $k-\text{triangles}$  for $k_1=0.61~\text{Mpc}^{-1}$ at $z=0.99$. \textbf{Bottom panel:} The signal-to-noise ratio of 21-cm auto bispectrum and 21-cm-galaxy cross-bispectrum for all unique $k-\text{triangles}$ for $k_1=0.61~\text{Mpc}^{-1}$ at $z=0.99$ with 200\,hours of observations for SKA-Mid AA4 for each pointing and Euclid-like galaxy redshift survey.}
    \label{fig:cross_bispec}
\end{figure}
To estimate the signal-to-noise (SNR) ratio of the 21-cm-galaxy cross-bispectrum, we simulate 50 statistically independent
realisations of the 21-cm system noise maps (for SKA-Mid observations) within the volume of the 21-cm signal, and add the
system noise and 21-cm signal. To estimate the 21-cm system noise map, we considered 200 hours of SKA-Mid AA4 observations for each pointing in the interferometric mode. For this forecast, we do not consider the uncertainty in the cross-bispectrum due to the cosmic variance. The upper panel of the \autoref{fig:cross_bispec} shows the 21-cm auto-bispectrum and 21-cm-galaxy cross-bispectrum for three different cross-combinations for all the unique $k-\text{triangles }$ for $k_1 = 0.61~\text{Mpc}^{-1}$ at redshift $z=0.99$. The SNR for the corresponding bispectra are shown in the bottom panel. The 21-cm-galaxy cross-bispectrum shows a significant enhancement in the SNR compared to the 21-cm auto-bispectrum, thanks to the higher SNR of the galaxy surveys. The cross-bispectra achieve SNR values above $5\sigma$ for all unique $k$-triangles, with linear triangles reaching SNRs greater than $10~\sigma$ \citep{Noble:2026oqm}.

\subsection{A Data-Driven Null Test for Multi-Line Intensity Mapping: The $\mathcal{Q}$ Estimator}
\label{sec:Q}

A central challenge in LIM is the difficulty in measuring unbiased auto-power spectra due to contamination from instrumental noise, residual foregrounds, systematics. Cross-correlating different lines has emerged as a robust alternative: cross-power spectra are less sensitive to uncorrelated contaminants and can infer physical quantities even when individual auto-spectra are inaccessible.

A formalism proposed by \cite{Beane:2018dzk} (hereafter \B19) demonstrated that, under certain assumptions, the auto-power spectrum of a line can be reconstructed solely from cross-power spectra with other lines. This requires that all lines involved (1) linearly trace the same underlying matter density field, and (2) are strongly mutually correlated. If these conditions are met, the auto-spectrum of line a can be estimated using
\begin{equation}
    \hat{P}_{aa} (k) = \frac{P_{ab}(k) P_{ac}(k)}{P_{bc}(k)}\,,
\end{equation}
where $P_{ab}$ is the cross-power spectrum between lines a and b, and so on.
While powerful, the \B19 method provides no internal consistency check: its reliability hinges on assumptions that may not hold across all spatial scales, redshifts, or line combinations. In particular, small-scale physics, scale-dependent bias, and the stochastic nature of different tracers can cause deviations from the idealized assumptions, limiting the scales over which cross-spectrum-based inference is trustworthy.

To address this limitation, we introduce a new diagnostic statistic, the $\mathcal{Q}$ estimator, which provides a data-driven null test for assessing the validity of the assumptions underpinning \B19. Defined as \citep{Sarkar:2025vml}
\begin{equation}
    \mathcal{Q}_{abcd}(k) = \frac{P_{ab}(k) \, P_{cd}(k)}{P_{ac}(k) \, P_{bd}(k)}\,,
\end{equation}
this quantity is constructed from four line-intensity fields. When all lines linearly trace the same underlying density field and are mutually well-correlated, $\mathcal{Q} \approx 1$. Deviations from unity signal a breakdown of one or both assumptions, effectively flagging regimes where auto-spectrum reconstruction via cross-correlations may become biased or unreliable. 

In post-reionization, HI is associated with the galaxies that are hosted by dark matter haloes.  Here 21-cm signal is positively correlated with the star formation lines. To test the estimator under realistic conditions, we first start by simulating the LIM fields. 
We consider three star formation lines, namely \CII\,$\mu{\rm m}$, 
\NII\,$\mu{\rm m}$ and \CI\,$\mu{\rm m}$, at $z=1$ along with the 21-cm line for this exploration. From now onward, for brevity, we'll drop $\mu{\rm m}$ from the star formation lines. We assume all the lines considered are hosted by dark matter halos.
We used halo catalogs from the Illustris TNG300 simulations at $z=1$. For the 21-cm signal, we used a semi-analytic prescription proposed in equation 13 in~\cite{Villaescusa-Navarro:2018vsg} to populate the halos with 21-cm signal. The values of the model parameters are taken from Table~1 in~\cite{Villaescusa-Navarro:2018vsg}.
In order to simulate the star formation lines, we used the publicly available \texttt{LIMpy} code~\citep{Roy:2023cpx}, which assigns line luminosities to halos through a two-step process. First, it computes the star formation rate (SFR) from halo mass $M_h$ using the fitting formula from~\cite{Fonseca:2016qqw},
\begin{equation}
\mathrm{SFR}(M_h, z) = M_0 \left( \frac{M_h}{M_1} \right)^\mathfrak{a} \left( 1 + \frac{M_h}{M_2} \right)^\mathfrak{b} \left( 1 + \frac{M_h}{M_3} \right)^\mathfrak{c} \,,
\end{equation}
where $M_1 = 10^8\,{\rm M}_\odot$, and $M_0, M_2, M_3, \mathfrak{a}, \mathfrak{b}, \mathfrak{c}$ are redshift-dependent parameters specified in Table 1 of~\cite{Fonseca:2016qqw}. Second, it converts the resulting SFR to line luminosity $L$ via the relation
\begin{equation}
\frac{L}{L_\odot} = 10^{\alpha_{\rm SFR}} \left( \frac{\mathrm{SFR}}{{\rm M}\odot,{\rm yr}^{-1}} \right)^{\beta_{\rm SFR}}\,.
\label{eq:luminosity_sfr}
\end{equation}
The coefficients $\alpha_{\rm SFR}$ and $\beta_{\rm SFR}$ differ across lines. 
For \CII\, \NII\ and \CI\, the ($\alpha_{\rm SFR}$,$\beta_{\rm SFR}$) pairs are taken to be 
(6.98,0.99), (5.70, 0.95) and (6.30, 0.50), respectively.
In addition, we impose a minimum halo mass threshold $M_{\rm min}=10^{9}\,{\rm M}_{\odot}$, below which halos are assumed incapable of hosting significant star formation and hence do not contribute to the line signal. Finally, we add realistic instrument noise to the simulated maps to mimic the real observations. 

\begin{figure}
    \centering
    \includegraphics[width=1\linewidth]{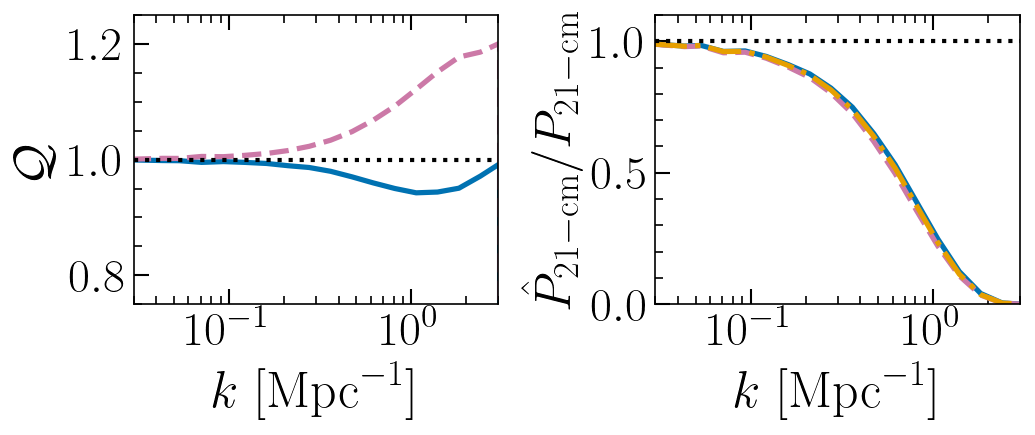}
    \caption{Diagnostics for evaluating the validity of the \B19 estimator at redshift \(z = 1\), using a set of four tracers: (A) 21-cm, (B) \CII\,$\mu$m, (C) \NII\,$\mu$m, and (D) \CI$\,\mu$m. The left panel shows the \(\Q\) estimator as a function of $k$ for two combinations of cross correlations, $(P_{\rm AB}P_{\rm CD})/(P_{\rm AC}P_{\rm BD})$ (solid sky-blue line) and $(P_{\rm AC}P_{\rm BD})/(P_{\rm AD}P_{\rm BC})$ (dashed magenta line).
    The right panel shows the \B19-estimated 21-cm power spectra \(\hat{P}_{\rm 21-cm}\) normalized by the true spectra \(P_{\rm 21-cm}\) as a function of $k$. The estimated 21-cm power spectra are shown for three combinations of cross correlations, \((P_{\rm AB}P_{\rm AC})/(P_{\rm BC})\) (solid sky-blue line), \((P_{\rm AB}P_{\rm AD})/(P_{\rm BD})\) (dashed sky-magenta line), \((P_{\rm AC}P_{\rm AD})/(P_{\rm CD})\) (dashed-dotted orange line).
}
    \label{fig:Q_no_noise}
\end{figure}

Before analyzing the estimator with noise, we assess the $\mathcal{Q}$ estimator without any instrument noise, that is pure signal. 
Figure~\ref{fig:Q_no_noise} shows the $\mathcal{Q}$ estimator as function of wave vector $k$, along with the \B19 estimator. We find that $\mathcal{Q}\approx1$ at $k\lesssim0.2\mpci$, where the \B19 estimator provides an unbiased estimation of the 21-cm power spectrum. At $k>0.2\mpci$ we see that $\mathcal{Q}$ deviates from $1$. The prediction of the \B19 estimator also departs from the true prediction beyond this $k$. This is possibly due to the fact that the non-linearities become significant at these $k$ values. This confirms that $\mathcal{Q}$ can be used as a diagnostic to assess the assumptions in the \B19 estimator and can pinpoint the range over which the estimates of the \B19 estimator can be trusted. We also found that $\mathcal{Q}$ is remarkably stable across a wide range of physically plausible models and scenarios.

\begin{figure}
    \centering
    \includegraphics[width=1\linewidth]{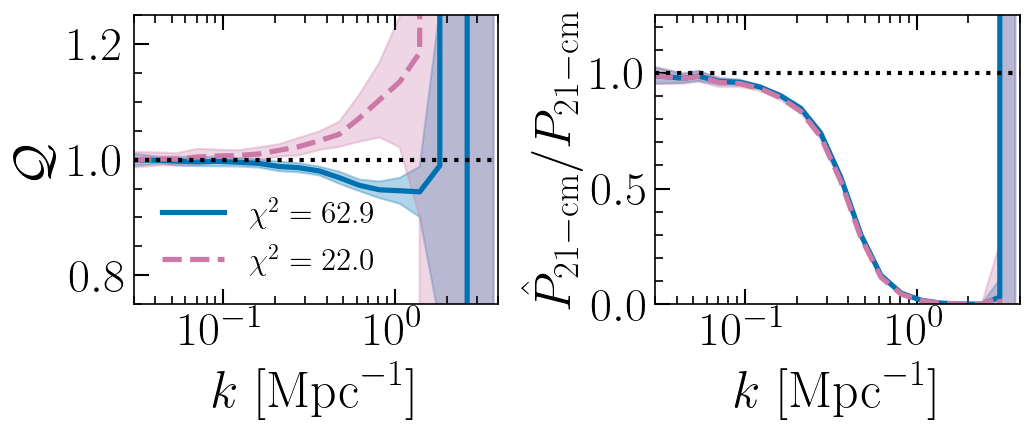}
    \caption{Same as Figure~\ref{fig:Q_no_noise} but including the effects of the instrument noise. The shaded regions show the $1\sigma$ scatter due to noise, estimated from 100
    independent noise realizations for each experiment. The $\chi^2$ score---computed as the squared, variance-weighted deviation from the true value (unity for $\mathcal{Q}$ and the true theoretical power spectrum for the \B19 estimator)---quantifies the significance with which any departure from the expected signal can be detected.
    }
    \label{fig:Q_with_noise}
\end{figure}

We next incorporate realistic instrumental noise for the corresponding LIM surveys and re-assess the performance of the $\mathcal{Q}$ estimator. For the 21-cm signal, we use the estimated noise for the SKA-Mid telescope, more precisely, the AA4 configuration. For the star formation lines, we assume a CCAT-prime-like single dish experiment with a slightly futuristic configuration. Our approach for modeling the noise closely follows~~\cite{Fronenberg:2024olu}, where we treat instrumental noise as a Gaussian random field with zero mean and root-mean-square (RMS) amplitude $\sigma_{\rm rms}$. For the 21-cm signal, $\sigma_{\rm rms}$ is computed following section 5.1 of~\cite{Fronenberg:2024olu}, where the SKA-Mid specific parameters are used from \href{https://www.skao.int/en/science-users/ska-tools/493/ska-sensitivity-calculators}{https://www.skao.int/en/science-users/ska-tools/493/ska-sensitivity-calculators}. 
The total observation time is taken to be $3000$ hrs.
For the star formation lines, the corresponding parameters for the $\sigma_{\rm rms}$ computation are taken from Table~2 of Ref.~\cite{Padmanabhan:2022elb}, where we specifically used the parameters for FYST-like (Stage II) experiment with one change, that is, we have considered $N_{\rm spec,eff}=1000$. The total observation time for each of the star formation lines is taken to be $1000$ hrs. After estimating $\sigma_{\rm rms}$ for the individual lines, we generate noise realizations to be added to the signal cubes. For each line, we used $100$ independent noise realizations and computed the \(Q\) and \B19 estimator for each. In Figure~\ref{fig:Q_with_noise}, we plot the mean and $1\sigma$ scatter of both the estimators computed using the $100$ realizations. We see that $\mathcal{Q}$ is still a reliable indicator of the success of the \B19 predictions even in the presence of realistic noise.

As upcoming LIM surveys increasingly leverage multi-line synergies, the $\mathcal{Q}$ diagnostic provides a critical tool for identifying the spatial and physical regimes in which joint cross-correlation techniques can be robustly applied.

\subsection{Marked Informed Cross spectrum}
\label{sec:mark}

The key idea behind Marked statistics~\citep{White_2016, Kamran:2024xob, Pandya:2026chj} is to weight the target LIM based on its local properties. Using this “mark” encodes additional information in the map and allows to calculate a more generalized two point correlation function. The great potential of multi line intensity mapping previously described in this chapter, motivated us to extend the cross power spectrum formalism introduced in the previous sections, generalizing it by using the idea of Marked statistics.
The idea is to weigh the various intensity maps based on some property of the line emission and to cross-correlate the resulting, more informed, maps. This hybrid statistic allows to combine the simple and versatile nature of the Marked statistic to capture the higher order information with the ability of the cross power spectrum to retain the phase difference between two distinct maps which can be used to carry out joint constraints on the astrophysical parameters governing two separate maps. 

To mark the field, we use the functional form inspired from \cite{White_2016, Kamran:2024xob}:

\begin{equation} \label{subsection MICS: m0}
    m_0(\mathbf{x},z;{\rm R},f,p) = m_0(\delta_{\rm R};f,p) = \Bigg[ \frac{1 + f}{2 + f + \delta_{\rm R}(\mathbf{x},z)} \Bigg]^p
\end{equation}
where $\delta_{\rm R}(\mathbf{x},z)$ smoothed fluctuation map with a spherical top hat function of radius ${\rm R}$. The combined choice of the free parameters, ${\rm R},f,p$ (the Mark parameters) decide of what property the target map is to be informed. For instance, the sign of the exponent decides if the over-densities or the under-densities, in the maps, are to be emphasized. The general idea of the Mark enables to extend beyond a simple functional form which can be modified to tackle the challenge at hand and the nature of the LIM too. After marking the maps, we cross-correlate the new fields as follows:

\begin{equation}\label{subsection MICS: MICS}
    \big\langle \mathbf{M}_{\rm a}(\bm{k}) \mathbf{M}^*_{\rm b}(\bm{k^\prime})\big\rangle = V \delta_{k,k^\prime} P_{\rm Ma\times Mb}(k)
\end{equation}

where $\mathbf{M}_{\rm a}(\bm{k})$ and $\mathbf{M}_{\rm b}(\bm{k})$ are the Marks on the fields $a$ and $b$ respectively in the Fourier domain and $V$ is the cosmological volume. The statistic $P_{\rm Ma\times Mb}(k)$ is the \textit{Marked Informed Cross spectrum (MICS)}. Analogous to the correlation coefficient of the cross power spectrum, we use the cross correlation coefficient to determine relative phase difference between the maps, which are additionally informed. This is defined as

\begin{equation}\label{subsection MICS: r_m0}
    r_{\rm Ma\times Mb}(k) = \frac{P_{\rm Ma\times Mb}(k)}{\sqrt{P_{\rm Ma}(k)\cdot P_{\rm Mb}(k)}}
\end{equation}
where $P_{\rm Ma}(k)$ and $P_{\rm Mb}(k)$ are the Marked power spectra of the maps $a$ and $b$. 

\begin{figure}
    \centering
    \includegraphics[width=1\linewidth]{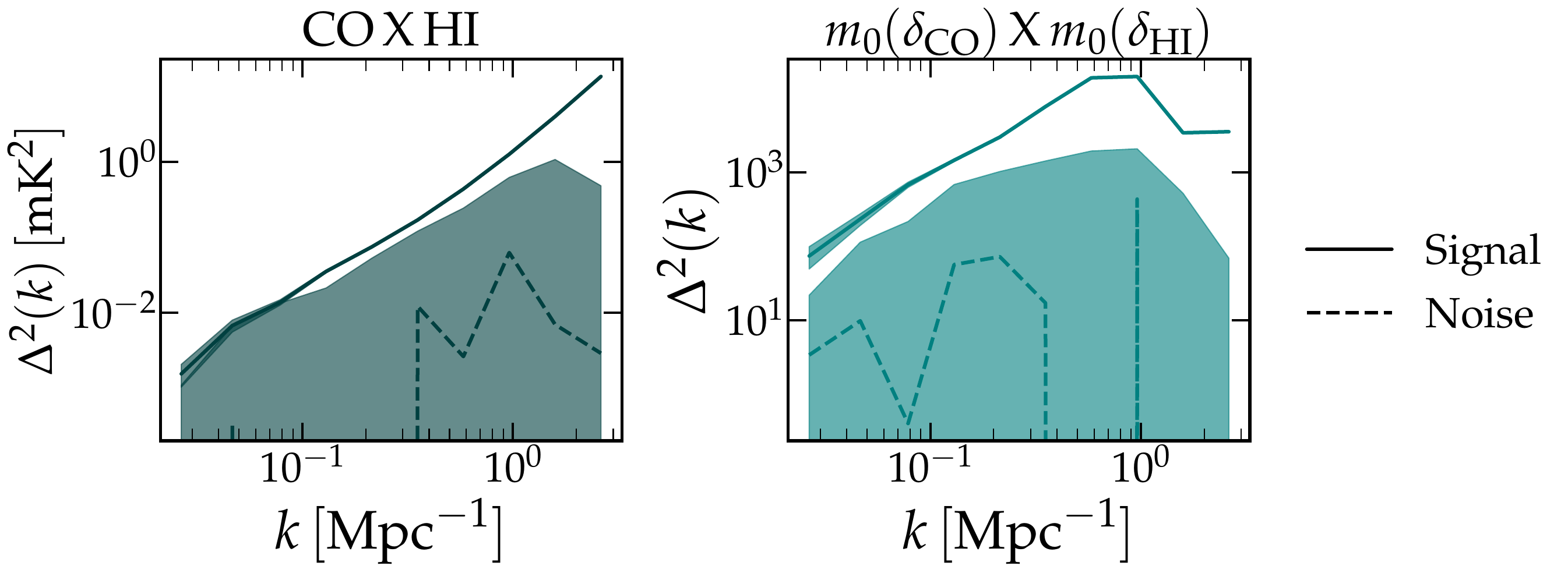}
    \caption{Comparing the signal strength, at different $k$-modes, against the corresponding noise levels using Marked statistics designed to boost the signal strength at redshift $z=1$ with the standard cross signal power spectrum. \textit{Left panel}-The dimensionless HI cross CO power spectrum of the signal and noise. \textit{Right panel}-The dimensionless HI cross CO Marked power spectrum of the signal and noise. The shaded region corresponds to $\pm1\sigma$ sigma spread around mean.}
    \label{fig:Cross_Mark_CO_HI_SNR.pdf}
\end{figure}

A suitable mark can extract the existing information in the map through a transformation, being an independent entity itself. The corresponding cross statistic can also be used to boost the signal strength, as the errors of different instruments, being random, for two distinct signals are uncorrelated. Moreover, the implementation of the mark can be generalized to highlight attributes in the field itself that may not be evident. We simulate the mock LIM noise maps, focusing on the [CO]–2.6\,mm and [HI] 21-cm emission lines at $z = 1$, using COMAP-like survey with 5000 hours of integration time and SKA-Mid AA4 configuration with 5000 hours of integration time. We mark the signal and the noise, by transforming the signal, similarly and estimate the cross power spectrum. Figure \ref{fig:Cross_Mark_CO_HI_SNR.pdf} shows it is principle possible to significantly boost the signal above noise levels.

\subsection{Other observational techniques}

A significant challenge in LIM modeling involves accurately characterizing the highly non-Gaussian information embedded within line-intensity maps \citep{2024Bernal}. Accessing this non-Gaussianity is crucial for differentiating between cosmological and astrophysical effects and for robust parameter inference. One-point statistics, such as the voxel intensity distribution (VID,~\cite{Breysse:2015saa}), provide direct access to this highly non-Gaussian information. Ongoing research is focused on developing analytical covariances between the VID and the LIM power spectrum to enhance the robustness of these statistical analyses \citep{Breysse:2022alx,2024Bernal}.

Furthermore, deep learning techniques are being explored to address some of the inherent complexities in LIM, such as disentangling contributions from different emission lines that overlap in observed wavelengths, a phenomenon known as spectral line de-confusion. Conditional generative adversarial networks (GANs) are a proposed method for tackling this spectral line de-confusion problem, offering a novel approach for component extraction and signal separation. Advanced simulations, such as those from the THESAN project~\citep{Kannan:2021ucy}, play a vital role in predicting multi-tracer LIM signals during the epoch of reionization, incorporating intricate astrophysical processes like photoionization and radiative transfer. Recent alternatives explored the possibilities of training conditional probability distributions on zoom-in simulations to then map the line luminosities over lower resolution, larger volume cosmological simulations~\citep{yang_2025}. Transformer-based generative frameworks are also been developed~\citep{Moriwaki:2025edv}.
These simulations provide crucial insights into how various spectral lines from the interstellar medium and 21-cm emission from neutral hydrogen gas can be utilized to study galaxy formation and evolution, as well as the fundamental cosmology of the early Universe \citep{2023Sato}.

\section{Morphological measures on multiline intensity map images}
\label{sec:morph}


Imaging line emission from galaxies across multiple spectral tracers in the intensity-mapping regime provides a powerful means to visualize the complex, web-like structure of the underlying matter distribution over a broad dynamic range.
The prospects for 21-cm--CO synergies with SKAO were first explored by \citet{Chang:2015era} at the EoR ($z\sim8$); here we extend this to the post-EoR regime using morphological statistics.
Beyond traditional Fourier statistics, non-Fourier image-based summaries can capture non-Gaussian and morphological information that is otherwise lost. These morphological measures can therefore provide additional constraints, help distinguish between astrophysical models, and break degeneracies in cosmological parameter inference.

\begin{figure}[htbp]
    \centering
    \includegraphics[width=0.55\linewidth]{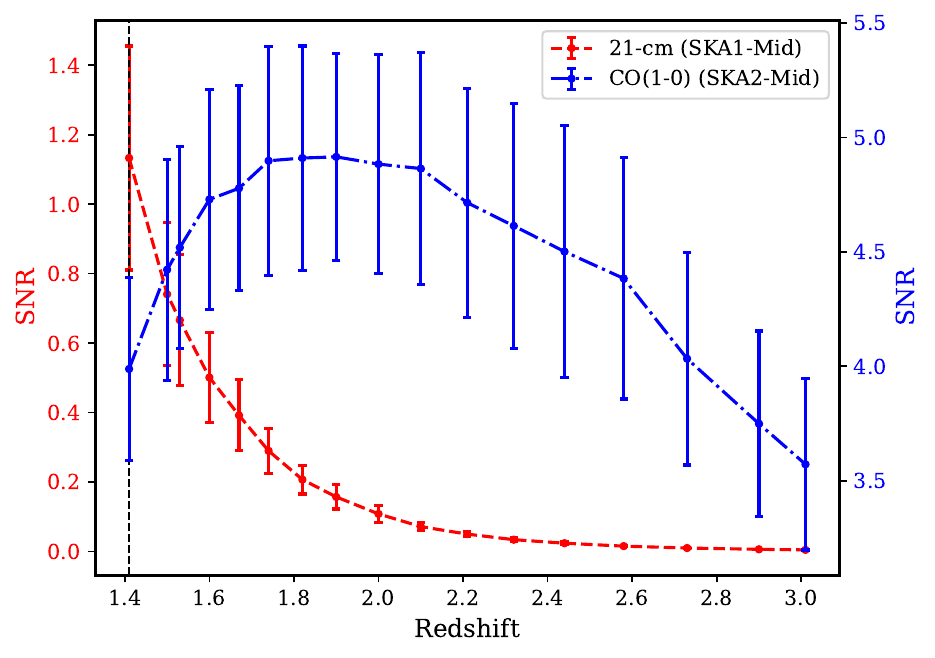}
    \caption{Signal-to-noise ratios (SNR) of the $21$-cm and CO($1-0$) brightness temperature fluctuations at spatial resolution $\delta x \simeq 1$ Mpc in the overlapping redshift range where SKA-Mid and SKA Phase 2 can observe the $21$-cm and CO($1-0$) LIM signals, respectively. The noise estimates are made assuming Gaussian random thermal noise for $5000$ hours per pointing of SKA-Mid observations and $100$ hours per pointing of SKA Phase 2 observations for the $21$-cm and CO($1-0$) signals, respectively. The black dashed line corresponds to $z=1.41$, the redshift where both the $21$-cm and CO($1-0$) brightness temperature fluctuations have a significant SNR. The $1\sigma$ error bars denote standard deviation in SNR across $8 \, (150 \, {\rm Mpc})^3$ 21-cm subcubes and $27 \, (100 \, {\rm Mpc})^3$ CO($1-0$) subcubes, corresponding to realistic SKA-Mid surveys, due to cosmic variance in signal fluctuations. See \cite{dosibhatla_2025_lss-morphology} for more details.}
    \label{fig:image_SNR}
\end{figure}

Figure~\ref{fig:image_SNR} shows the signal-to-noise ratio (SNR) estimates for mock 21-cm maps observed with SKA-Mid and CO(1--0) maps observed with the proposed SKA Phase 2 Band~6\footnote{\href{https://www.skao.int/sites/default/files/documents/d38-ScienceCase_band6_Feb2020.pdf}{SKAO
 Technical Document on Band~6; see Section~6.7.}} \citep{dosibhatla_2025_lss-morphology}. Both sets of maps are generated at the same redshifts and include only Gaussian thermal noise. The signal fields are obtained by post-processing subhalo catalogues from the IllustrisTNG simulation following the methodology of \citet{dosibhatla_2025_lss-morphology}.

Constructing high-resolution intensity maps is challenging due to bright foregrounds and instrumental systematics, both of which exceed the intrinsic signal by several orders of magnitude. Thermal noise further erases the small-scale information. Achieving high angular resolution requires long interferometric baselines, which are sparsely sampled, while high line-of-sight resolution demands narrow frequency channels that increase noise variance.


Even with 5000~h of integration per pointing and a spectral resolution corresponding to eight SKA-Mid channels, the 21-cm maps remain noise dominated, except near $z \approx 1.4$ where the SNR approaches unity. By contrast, the CO(1--0) maps remain signal-dominated even with only 100~h of observing time per pointing. Motivated by this behaviour, the image analysis results presented in this chapter focus on $z=1.41$, where both tracers exhibit significant SNR.

\subsection{Largest Cluster Statistic}
\label{subsubsec:LCS_formalism}

In a binary field that can be segmented into clusters and regions not belonging to any cluster, the largest cluster statistic (LCS) is defined as the fraction of cluster volume lying in the largest cluster \citep{klypin_1993_percolation, Bharadwaj_2000, BagMondal_2018, Pathak_2022, Dasgupta_2023, pal_2025_lcs}:
\begin{equation}
    {\rm LCS} = \frac{\text{Volume of the largest cluster}}{\text{Total cluster volume}} \, .
\end{equation}
We define a cluster as a connected set of bright cells in an intensity map.

The percolation transition refers to the onset of large-scale connectivity in the cosmic web, where the matter distribution connects to form a single, contiguous, large cluster spanning the entire universe. The LCS is useful for studying the percolation transition of an intensity map \citep{shandarin_1983_percolation, klypin_1993_percolation, Bharadwaj_2000, Pathak_2022, Dasgupta_2023, Regos_2024, pal_2025_lcs}, which is quantified by the evolution of the LCS with the filling factor (FF) of the map, defined as
\begin{equation}
    {\rm FF} = \frac{\text{Total cluster volume}}{\text{Total survey volume}} \, .
\end{equation}

The total cluster volume, and hence the filling factor, grow steadily with the growth of structure. However, at the stage where large clusters start to merge, the volume of the largest cluster grows sharply and becomes comparable to the total cluster volume, making the LCS approach $1$. This marks the percolation transition in an intensity map.

We employ an iterative coarse-graining scheme as in \citep{Bharadwaj_2000, dosibhatla_2025_lss-morphology} and estimate the LCS and FF at every iteration. The percolation curve is obtained, which tracks the evolution of the LCS with FF. The percolation transition is attained at different filling factors for different intensity maps, allowing us to distinguish between their morphologies.

\begin{figure}[htbp]
    \centering
    \includegraphics[width=0.7\textwidth]{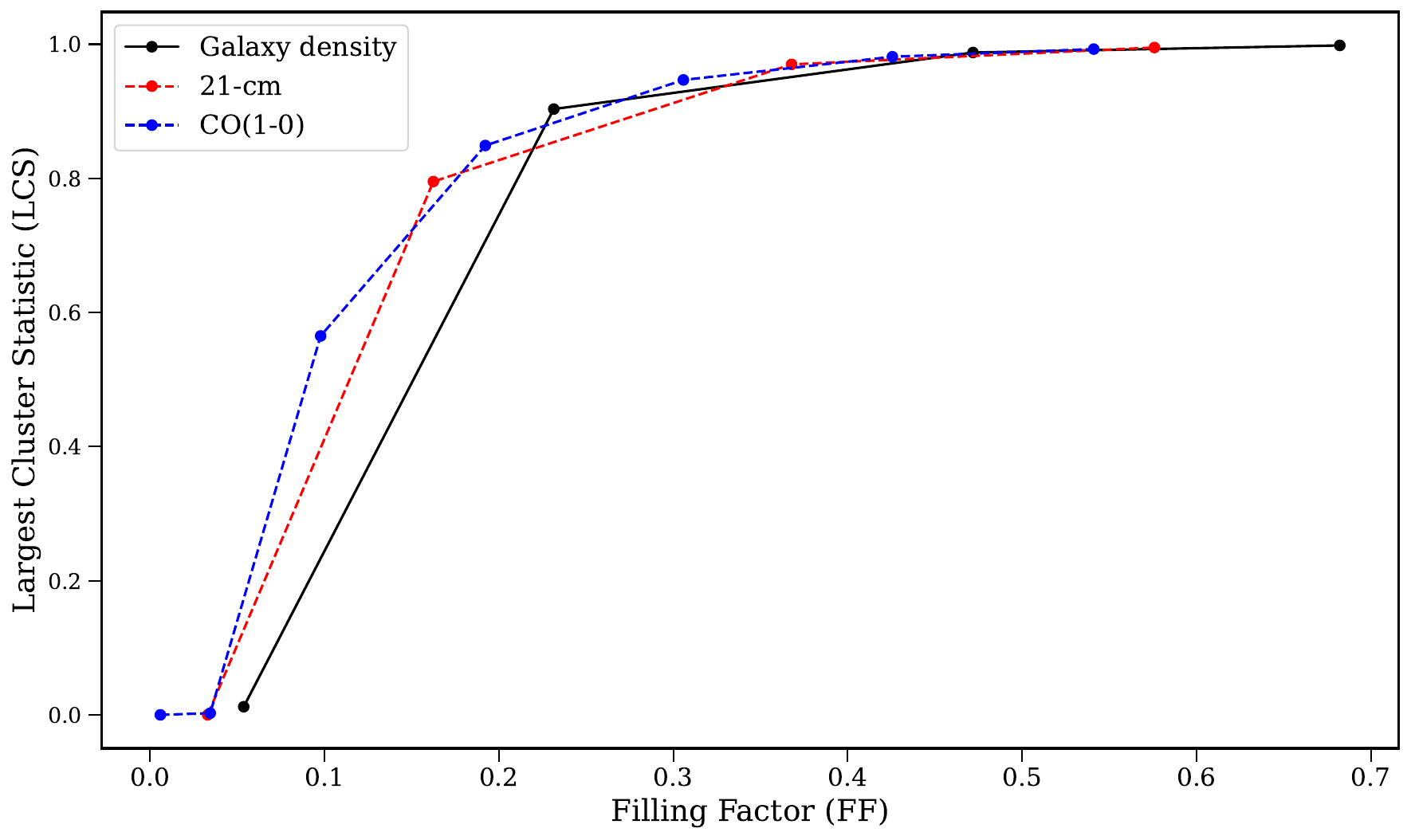}
    \caption{Percolation curves for the galaxy mass density (solid black) maps and 21-cm (dashed red) and CO(1--0) brightness temperature (dashed blue) maps at a spatial resolution of $0.5$ Mpc. See \cite{dosibhatla_2025_lss-morphology} for more details.}
    \label{fig:percolation}
\end{figure}

Figure \ref{fig:percolation} shows the percolation curves for the galaxy density field and 21-cm and CO(1--0) intensity maps at $z=1.41$, simulated from the same underlying galaxy distribution. The FF and LCS values are computed after every iteration of the coarse-graining scheme mentioned in section \ref{subsubsec:LCS_formalism}. The three maps follow different percolation curves, despite tracing the same galaxy distribution. The LCS is therefore sensitive to the complex astrophysics of line emissions, such as the star formation rate and cold gas content. Due to its sensitivity to factors that affect the connectivity and filamentarity of the LSS, LCS can be an effective higher-order summary statistic to infer both astrophysics and cosmology.
 
\subsection{Local Dimension}
\label{subsubsec:local_dimension_formalism}

The LSS of the universe consists of filaments, sheets, nodes, and voids, which form a cellular structure known as the ``cosmic web" \citep{gregory_1978_cosmic-web, einasto_1984_cosmic-web, zeldovich_1982_giant-voids, bond_1996_cosmic-web, Bharadwaj_2000, aragon-calvo_2010_cosmic-web, libeskind_2017_cosmic-web, tojeiro_2025_cosmic-web_review}. The contribution of the different building blocks of the cosmic web to the line emission in an intensity map contains a wealth of non-Gaussian and morphological information. The local dimension \citep{sarkar_2009_locdim, sarkar_2012_locdim, sarkar_2019_locdim, pandey_2020_locdim} is a simple yet effective way of determining whether a cell in an intensity map belongs to a filament, sheet, or a volume-filling environment (node or void) \citep{dosibhatla_2025_lss-morphology}. The local dimension ($D$) of a cell in a map determines how the number of bright cells ($N(R)$) in a sphere centered at the given bright cell scales with the radius of the sphere.
\begin{equation}
    N(R) = AR^D ,
    \label{eq:local_dimension}
\end{equation}
where $A$ is a normalisation constant. For straight filaments and plane sheets passing through the centre of the sphere, and volume-filling environments, $D$ takes values $\simeq$ $1$, $2$, and $3$ respectively. However, fractional values are allowed for environments intermediate between any two kinds.

To compute the local dimension of any bright cell in an intensity map, we count $N(R)$ by incrementing $R$ by a multiple $\Delta R$ of the grid resolution between fixed values $R_{\rm min}$ and $R_{\rm max}$. We then fit the variation of $N(R)$ with $R$ using equation \ref{eq:local_dimension}. If the fit converges with $\chi^2_\nu \leq \chi^2_{\nu,{\rm max}}$, the cell is classifiable, and it belongs to the environment corresponding to $D$ in the length scale $R_{\rm min}$ to $R_{\rm max}$. The value of $\chi^2_{\nu,{\rm max}}$ is chosen such that a good fit is obtained and a large number of cells is classifiable. The fraction of cells in different environments effectively characterizes the morphology of an intensity map.

\begin{figure}[htbp]
    \centering
    \includegraphics[width=0.6\linewidth]{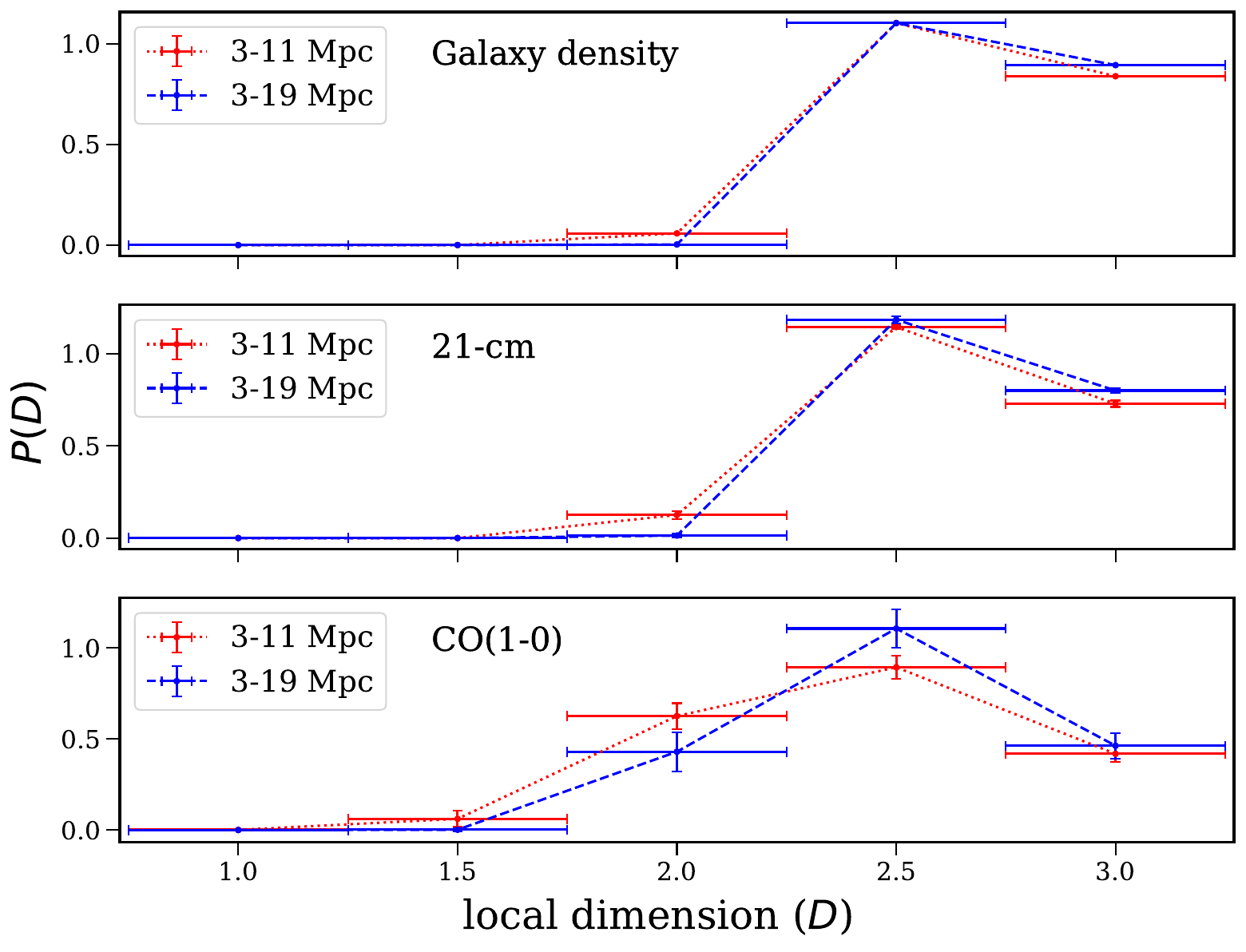}
    \caption{Distribution of local dimensions of classifiable cells out of $10^5$ randomly sampled bright cells in the galaxy mass density (top), $21$-cm (middle), and CO($1-0$) (bottom) maps simulated with galaxy positions in redshift space. The $D$-values are binned into intervals of size 0.5. The different curves correspond to different length scales specified by the $R_{\rm min}$ and $R_{\rm max}$ values. The vertical $1\sigma$ error bars denote standard deviation in $P(D)$ across $8 \, (150 \, {\rm Mpc})^3$ 21-cm subcubes and $27 \, (100 \, {\rm Mpc})^3$ CO($1-0$) subcubes, corresponding to realistic SKA-Mid survey areas of $4 \, {\rm deg}^2$ and $1.77 \, {\rm deg}^2$ respectively for $10^4$ randomly sampled bright cells. See \cite{dosibhatla_2025_lss-morphology} for more details.}
    \label{fig:locdim_signal_RSD}
\end{figure}

Once the local dimensions of a subset of cells in an intensity map are determined, the fraction of cells lying in filaments, sheets, and volume-filling environments (nodes and voids) can be computed. Figure \ref{fig:locdim_signal_RSD} shows the distribution of local dimensions of the galaxy density field and 21-cm and CO(1--0) intensity maps divided into bins of width $0.5$. The maps have a spatial resolution of $\delta s = 0.87$ Mpc, corresponding to a spectral resolution of $\delta \nu = 107.52$ kHz (channel-averaged spectral resolution of eight SKA-Mid channels).

The 21-cm and galaxy density maps exhibit a similar morphology, with regions intermediate between sheets and voids accounting for most of the galaxy overdensity and 21-cm emission. However, the CO(1--0) emission from void-like regions is suppressed. The star formation rates of galaxies are the main distinguishing factor between the morphologies of the 21-cm and CO(1--0) maps. Galaxies that emit the 21-cm line but have very low or no CO(1--0) luminosity are the ones where the star formation is quenched. It is therefore found that the local dimension can be used to distinguish between such different astrophysical scenarios. The detectability of the local dimensions is studied in the presence of thermal noise and line interlopers in \cite{dosibhatla_2025_lss-morphology}.

\section{Discussion}
\label{sec:discussion}

\subsection{Prospects and key takeaways}

SKAO-enabled multi-line intensity mapping offers several transformative opportunities for cosmology and galaxy evolution. Cross-correlations of the 21-cm signal with tracers such as [CII], CO, H$\alpha$, or Ly$\alpha$ substantially reduce the impact of uncorrelated systematics, providing a robust path to early high-significance detections of 21-cm clustering. Measurements of RSD multipoles in cross-correlations can extract $f\sigma_8$ and Alcock–Paczynski distortions with minimal reliance on potentially biased auto-spectra.

At the same time, combining 21-cm emission with molecular and PDR tracers directly probes the baryon cycle by linking atomic gas, molecular gas, and star-forming regions within the same halos. One-halo and shot-noise contributions help constrain duty cycles and halo occupation. Beyond two-point statistics, cross-bispectra and marked cross-correlations provide access to non-linear bias and higher-order information that cannot be captured by power spectra alone \citep{Majumdar01.2026.SKA}. The $\mathcal{Q}$ estimator further offers a practical data-driven test of the assumptions underlying cross-only auto reconstructions (\B19), identifying the scales where cross-based inference remains reliable.

\subsection{Analysis challenges}

These opportunities come with practical challenges. Differences in beams, bandpasses, and masks complicate the direct comparison of modes across surveys, requiring careful regridding and accurate instrument modeling. Although cross-correlations suppress many contaminants, correlated interlopers and spatially correlated continuum emission can still leak into the signal, making tomographic checks and anisotropy tests essential.

Degeneracies among mean intensities, biases, and duty cycles persist in two-point analyses, and must be reduced using small-scale one-halo information, cross-bispectra, or priors from resolved surveys such as ALMA or JWST. Calibration stability is also critical: bandpass variations, polarization leakage, and relative gain drifts can bias even cross-power spectra. Null tests—including $\mathcal{Q}$, phase randomization, and time-split consistency—are therefore required. Finally, theoretical uncertainties limit the usable scale range; quasi-linear or EFT-inspired templates and simulation-calibrated emulators help maintain robustness when defining the smallest scales that can be used.

\subsection{A practical roadmap}

A practical strategy for exploiting multi-line LIM synergies begins with survey design that prioritizes sky and redshift overlap between SKAO Band~1 observations and surveys targeting [CII] or CO, while ensuring that spectral resolutions are sufficiently compatible to preserve common support in $k_\parallel$. Early detections should rely primarily on cross-power spectra—particularly their multipoles and anisotropic wedges—using pipelines explicitly designed to avoid noise-biased auto-spectrum estimates.

The $\mathcal{Q}$ estimator plays a central role in determining the reliable scale range for such analyses. Its behaviour as a function of $k$ identifies the wavenumbers for which the \B19 estimator provides unbiased reconstructions of auto-spectra and for which simple linear-bias modeling remains adequate for cosmological inference. With these scales in hand, one can progressively incorporate additional cross-information. Small-scale one-halo and shot-noise contributions, as well as marked cross-correlations, sharpen astrophysical constraints while maintaining conservative cuts for cosmology. High-SNR cross-bispectrum configurations—especially squeezed and linear shapes—provide further sensitivity to second-order bias parameters and stochasticity, and can be integrated into joint likelihood frameworks. Throughout, a rigorous robustness suite is essential, including time-split tests, interloper-induced anisotropy checks, transfer-function recovery from end-to-end simulations, and marginalization over calibration uncertainties.

\subsection{Implications for SKAO}
By anchoring a multi-tracer LIM program, SKAO can deliver competitive growth and distance constraints at $0.3\!\lesssim\!z\!\lesssim\!3$ \emph{and} uniquely connect them to gas-physics parameters that underlie star formation. 
With coordinated overlap to FIR/mm and optical/near-IR LIM, SKAO can provide the backbone for cross-correlation cosmology that is robust to many of the systematics that challenge single-experiment analyses.

\section*{Author List Ordering}
Authors for this chapter are ordered according to their overall level of contribution, in line with that expected for a small author list publication.

\section*{Acknowledgment}
DS acknowledges the support of the Canada 150 Chairs program, the Fonds de recherche du Québec Nature et Technologies (FRQNT) and the Natural Sciences and Engineering Research Council of Canada (NSERC) joint NOVA grant, and the Trottier Space Institute Postdoctoral Fellowship program.
SM, LN, MMD and AD acknowledge financial support through the Core Research Grant titled ``Observing the Cosmic Dawn in Multicolour using Next Generation Telescopes'' from the Science and Engineering Research Board (SERB) and the Department of Science and Technology (DST), Government of India. SM, MV, AD, LN, SKP and MMD acknowledge financial support through the project titled “Illuminating the Dark Sector of the Cosmos in the SKA Era” (Project No. P3497) funded under the “Scheme for Promotion of Academic and Research Collaboration (SPARC)” from the Ministry of Education, India. LN acknowledge the support from the Abdus Salam International Centre for Theoretical Physics (ICTP) under the `ICTP Sandwich Training Educational Programme (STEP)’  SMR.3991 and SMR.4129.   SKP, LN and YM acknowledge the financial support by the Department of Science and Technology, Government of India, through the INSPIRE Fellowship. JLB acknowledges funding from the grant UC-LIME (PID2022-140670NA-I00), financed by MCIN/AEI/ 10.13039/501100011033/FEDER, UE. AKS acknowledges the support from the National Science Foundation (grant No. 2206602). CH's work is funded by the Volkswagen Foundation, and supported through
Germany's Excellence Strategy through EXC~2181/1 -- 390900948 (the
Heidelberg STRUCTURES Excellence Cluster) and with support by the
Federal Ministry of Education and Research (BMBF) and the Ministry of
Science, Research and the Arts of Baden-Württemberg.

\bibliographystyle{abbrvnat-maxbibnames4}
\bibliography{chapter} 

\appendix




\end{document}

%% file: journal-names.tex
\newcommand{\actaa}{Acta Astron.} 
\newcommand{\araa}{Annu. Rev. Astron. Astrophys.} 
\newcommand{\aar}{Astron. Astrophys. Rev.} 
\newcommand{\ab}{Astrobiol.} 
\newcommand{\aj}{Astron. J.} 
\newcommand{\apj}{Astrophys. J.} 
\newcommand{\apjl}{Astrophys. J. Lett.} 
\newcommand{\apjs}{Astrophys. J. Suppl. Ser.} 
\newcommand{\ao}{Appl. Opt.} 
\newcommand{\apss}{Astrophys. Space Sci.} 
\newcommand{\aap}{Astron. Astrophys.} 
\newcommand{\aapr}{Astron. Astrophys. Rev.} 
\newcommand{\aaps}{Astron. Astrophys. Suppl.} 
\newcommand{\baas}{Bull. Am. Astron. Soc.} 
\newcommand{\caa}{Chinese Astron. Astrophys.} 
\newcommand{\cjaa}{Chinese J. Astron. Astrophys.} 
\newcommand{\cqg}{Class. Quantum Gravity} 
\newcommand{\gal}{Galaxies} 
\newcommand{\gca}{Geochim. Cosmochim. Acta} 
\newcommand{\icarus}{Icarus} 
\newcommand{\jcap}{J. Cosmol. Astropart. Phys.} 
\newcommand{\jgr}{J. Geophys. Res.} 
\newcommand{\jgrp}{J. Geophys. Res.: Planets} 
\newcommand{\jqsrt}{J. Quant. Spectrosc. Radiat. Transf.} 
\newcommand{\memsai}{Mem. Soc. Astron. Italiana} 
\newcommand{\mnras}{Mon. Not. R. Astron. Soc.} 
\newcommand{\nat}{Nature} 
\newcommand{\nastro}{Nat. Astron.} 
\newcommand{\ncomms}{Nat. Commun.} 
\newcommand{\nphys}{Nat. Phys.} 
\newcommand{\na}{New Astron.} 
\newcommand{\nar}{New Astron. Rev.} 
\newcommand{\physrep}{Phys. Rep.} 
\newcommand{\pra}{Phys. Rev. A} 
\newcommand{\prb}{Phys. Rev. B} 
\newcommand{\prc}{Phys. Rev. C} 
\newcommand{\prd}{Phys. Rev. D} 
\newcommand{\pre}{Phys. Rev. E} 
\newcommand{\prl}{Phys. Rev. Lett.} 
\newcommand{\psj}{Planet. Sci. J.} 
\newcommand{\planss}{Planet. Space Sci.} 
\newcommand{\pnas}{Proc. Natl Acad. Sci. USA} 
\newcommand{\procspie}{Proc. SPIE} 
\newcommand{\pasa}{Publ. Astron. Soc. Aust.} 
\newcommand{\pasj}{Publ. Astron. Soc. Jpn} 
\newcommand{\pasp}{Publ. Astron. Soc. Pac.} 
\newcommand{\rmxaa}{Rev. Mexicana Astron. Astrofis.} 
\newcommand{\sci}{Science} 
\newcommand{\sciadv}{Sci. Adv.} 
\newcommand{\solphys}{Sol. Phys.} 
\newcommand{\sovast}{Soviet Ast.} 
\newcommand{\ssr}{Space Sci. Rev.} 
\newcommand{\uni}{Universe} 